\documentclass[showpacs,amssymb,floatfix,pre,aps,notitlepage]{revtex4-1}

\usepackage{dcolumn}   
\usepackage{bm}        
\usepackage{amssymb}   

\usepackage{amssymb,amsfonts,bm}
\usepackage{graphics,graphicx}
\usepackage{colordvi,amsmath,epsfig,color}
\usepackage[activeacute,english]{babel}
\newcommand{\sig}{\boldsymbol{\widehat{\sigma}}}
\providecommand{\abs}[1]{\lvert#1\rvert}

\begin{document}

\title{Fate of Boltzmann's breathers:  kinetic theory perspective}

\author{P. Maynar}
\affiliation{F\'isica Te\'orica, Universidad de Sevilla, Apartado de Correos 1065, E-41080, Sevilla, Spain}
\author{M.I. Garc\'ia de Soria}
\affiliation{F\'isica Te\'orica, Universidad de Sevilla, Apartado de Correos 1065, E-41080, Sevilla, Spain}
\author{David Gu\'ery-Odelin}
\affiliation{Laboratoire Collisions, Agr\'egats, R\'eactivit\'e, FeRMI, Université  Toulouse III - Paul Sabatier, 118 Route de Narbonne, 31062 Toulouse Cedex 09, France}
\author{Emmanuel Trizac}
\affiliation{LPTMS, UMR 8626, CNRS, Universit\'e Paris-Saclay, 91405 Orsay, France}
\affiliation{ENS de Lyon, F-69364 Lyon, France}

\date{\today}

\begin{abstract}
The dynamics of a system composed of elastic hard particles confined
by an isotropic harmonic potential are studied. In the low-density
limit, the Boltzmann equation provides an excellent description, and the
system does not reach equilibrium except for highly specific
initial conditions: it generically evolves towards and stays in a breathing mode.
This state is periodic in time, 
with a Gaussian velocity distribution, an oscillating temperature
and a  density profile that oscillates as well. We characterize this breather
in terms of initial conditions, and constants of the motion.
For low but finite 
densities, the analysis requires to take into account 
the finite size of the particles. Under well-controlled
approximations, a closed description is provided, 
which shows how equilibrium is reached at long times. 
The (weak) dissipation at work erodes the breather's amplitude,
while concomitantly shifting its oscillation frequency. 
An excellent agreement is found
between Molecular Dynamics simulation results and the theoretical
predictions for the frequency shift. For the damping time, the agreement is not as accurate as for the frequency and the origin of the discrepancies is discussed.  
\end{abstract}

\maketitle

\section{Introduction}

The Boltzmann equation determines the dynamics of the one-particle
distribution function, $f(\bm{r},\bm{v},t)$, of a dilute gas
composed of particles with  short-range interaction 
($\bm{r}$ denoting position, $\bm{v}$ velocity, and $t$ time). The system can be confined by an external force \cite{boltzmann, resibois,
  cercignani}. Essentially, the dynamics of the distribution function  
can be decomposed into streaming due to the external force and binary
collisions between the particles. The equation was derived by
Boltzmann by ``counting'' collisions under the assumption of molecular
chaos, i.e. absence of correlations between the particles that are
going to collide. Nearly a century later, the equation could be
derived in a rigorous mathematical way for hard core systems
\cite{landford, landfordLibro}: assuming uncorrelated initial conditions, it can be proved in the
low density limit that the equation holds in some time window  \cite{nota1}. Nowadays, the Boltzmann equation is
an active area of research in Mathematics \cite{villani,
  raymond}. 
Besides, it is one of the
cornerstones of Statistical Physics \cite{resibois, cercignani}. It is widely used to study transport phenomena in different areas (solid state, plasmas, granular systems, traffic flow, population dynamics). To pinpoint just one of many applications: it is at the root of the lattice Boltzmann method, that has gained a prominent role as a key computational tool for a wide variety of complex states of flowing matter across a broad range of scales, from turbulence, to biosystems or nanofluidics and recently to quantum-relativistic subnuclear fluids \cite{succi}.

Apart from the fact that the state of the system can be described in
terms of the one-particle distribution function with a known evolution
equation, Boltzmann introduced a functional, 
$H[f]\equiv\int d\bm{r}\int
d\bm{v}f(\bm{r},\bm{v},t)\log f(\bm{r},\bm{v},t)$ 
from which the irreversible dynamics of the gas can be understood. 
More precisely, he proved the so-called $H$-theorem that states that, if
$f(\bm{r},\bm{v},t)$ satisfies the Boltzmann equation, then
$\frac{d H[f]}{dt}\le 0$. As $H$ is bounded from below, it must reach
a stationary value and this only occurs when $\log f$ is a collisional
invariant or, equivalently, when the distribution function is a Maxwellian
with some space and time dependent density, flow velocity and
temperature.  
By introducing this distribution into the Boltzmann equation, it can
be seen that, except for some external potentials that will be
considered later, the only possibility is to have spatial
and time independent temperature and flow velocity, with a density
field given by the equilibrium one. Then, invoking the above argument,
it can be proved that independently of the initial condition, the equilibrium
Maxwell-Boltzmann distribution is reached in the long-time
limit. Yet, Boltzmann himself realized that if the system is confined by
an isotropic harmonic potential, there exist more exotic solutions in
which the distribution function is a local equilibrium
Maxwell-Boltzmann distribution, but with some specific space and
time-dependent hydrodynamic fields \cite{b1909}: the
temperature is spatially homogeneous, the 
flow velocity is linear in $\bm{r}$ and the density is Gaussian.
The temperature and the cloud size oscillate with the same
frequency but with opposite phases around their equilibrium values. The
amplitude of the flow velocity also oscillates with the same frequency,
vanishing when the temperature is extremum. 
These ``breathing modes'' result from a perpetual conversion of kinetic and
potential energy, through a swing-like mechanism. 
These modes also exist beyond the harmonic confinement, for specific potentials worked out in \cite{gmrt14,DGO_2023}.

The breathing modes have long been considered as a mathematical curiosity
without any possibility of realization in Nature \cite{u63}. 
Any small imperfection in the trapping potential 
will eventually thermalize the system,
and ``pin'' it at equilibrium. In
the last years however, the interest for these modes has increased, specially in the
context of cold atoms where the system is typically confined by a
magnetic trap and the resulting 
confining potential can be accurately considered to be
harmonic \cite{dgps99}. Moreover, if the temperature
is somewhat larger than the critical temperature for the Bose-Einstein
condensate, the dynamics of the system are well described by the
classical Boltzmann equation. When the confined harmonic potential is
anisotropic, there still exist breathing modes but damped and with a
different frequency of oscillations with respect to the isotropic case. 
The frequency and damping coefficient of some modes (the
monopole and quadrupole ones) can be calculated from the Boltzmann
equation \cite{gzds99}, yielding a very good agreement with experiments
\cite{bpkw05} in the above mentioned conditions. Remarkably,
recently, thanks to a  
new magnetic trap capable of producing near-isotropic harmonic
potential, the non-damping monopole (i.e. the breather) has been experimentally
observed \cite{lbcl2015, GOT15}. Actually, the mode decays very slowly as a
consequence of small anharmonic perturbations, which decrease with
cloud size. Beyond a mere mathematical
curiosity, 
the relevance of the breather  is highlighted in Ref. \cite{lbcl2015}, as its relaxation
time can be made as small as desired with respect to all the
other modes. This leaves open the question of a possible thermalization of the system, in a perfectly isotropic and harmonic trap.

In this paper, we consider a system confined by an isotropic harmonic
potential with a twofold objective. The first is to 
clarify, at the level of the Boltzmann equation, which type of breather emerges out of the interactions. 
Second, the goal is to analyze 
the problem beyond the Boltzmann level, in a finite (although small) density system. 
To this end, we 
will consider that the particles are hard particles and the questions of interest are: do breathers survive and if so, how are they affected?
Is equilibrium reached in the long time limit? 
We will see that a breather indeed appears, which oscillates with a frequency that is slightly shifted
with respect to its Boltzmann counterpart, and that its amplitude 
decays very slowly in time. Close to equilibrium, we will compute the
frequency shift of the oscillations as well as the damping time. The same questions were recently tackled in \cite{gmgt24} using a hydrodynamic description, finding similar results (but restricted, by construction, to the hydrodynamic regime). It was found that the damping is associated to the nonlocality of collisions (meaning the fact that the centers of mass of particles colliding are at slightly different locations), so that it vanishes in the low density limit. Remarkably, for finite densities, the equilibrium relaxation time is related to the usually neglected bulk viscosity. This is so because the relaxation is carried out at constant temperature and with a shearless velocity field. In the present paper, it is shown that the results of \cite{gmgt24} are also valid beyond the hydrodynamic scale. In addition, by performing the analysis microscopically, an intuitive understanding of the damping mechanism at the particle level is achieved. 

The paper is organized as follows. The model is
introduced in section \ref{secModel}.
In
Sec. \ref{sec3}, the dynamics of the system are studied in the low
density limit. The main properties of the breathers are derived, and
the theoretical predictions are compared with Molecular Dynamics (MD)
simulations; the agreement is very good. In
Sec. \ref{sec4}, the dynamics of the system are studied beyond the
Boltzmann limit, 
accounting for
asymptotically long times (not accessible otherwise). 
A closed description is obtained, featuring a novel dissipation mechanism, impinging of the breather's late evolution.
It is found that the 
amplitude of their oscillations decays, and that the corresponding frequency is shifted with respect to the Boltzmann limiting case. 
Testing the predictions against MD simulations yields an excellent agreement for the
frequency;  the damping time is not as accurately captured, but its
scaling behaviour is found to match the predicted one. 
A discussion is presented in section \ref{sec:concl}, together with our conclusions.

\section{The model}
\label{secModel}
We consider $N$ elastic hard particles of mass $m$
and diameter $\sigma$. Let $\bm{r}_i(t)$ and $\bm{v}_i(t)$
denote the position and velocity of particle $i$ at time
$t$, respectively, with $i=1,\dots, N$. The system is confined by an isotropic harmonic potential: a particle at position $\bm{r}_i$ is subject to a force $\bm{F}_i=-k\bm{r}_i$, where $k>0$ is the stiffness. 
The spatial dimension of the system is $d$ ($d=2$ disks, $d=3$ for spheres). Upon
binary encounter between two particles, say particle $i$ and $j$, of
velocities $\bm{v}_i$ and $\bm{v}_j$, the velocities are
instantaneously changed to the postcollisional values, 
$\bm{v}_i'$ and $\bm{v}_j'$, in such a way that total momentum
and energy are conserved. 
\begin{eqnarray}
\bm{v}_i'
&=&\bm{v}_i+(\bm{v}_{ji}\cdot\sig)\sig, \label{cr1}\\
\bm{v}_j'
&=&\bm{v}_j-(\bm{v}_{ji}\cdot\sig)\sig, \label{cr2}
\end{eqnarray}
where $\bm{v}_{ji}\equiv\bm{v}_j-\bm{v}_i$ is the relative
velocity and $\sig$ a unit vector joining 
the two centers of particles at contact.

As the collisions are elastic, the total energy
\begin{equation}
E=\frac{m}{2}\sum_{i=1}^N[v_i^2(t)+\omega^2r_i^2(t)], 
\end{equation}
is conserved. Here, the  characteristic frequency of the
oscillations of one particle, $\omega \equiv\sqrt{\frac{k}{m}}$, has been
introduced. In addition, the total angular momentum
\begin{equation}
\bm{L}=m\sum_{i=1}^N\bm{r}_i(t)\times\bm{v}_i(t), 
\end{equation}
relative to the center of the force is also a constant of the
motion. On the other hand, although momentum is conserved in the
instantaneous collisions, total momentum is not a constant of the
motion due to the presence of the external force. 

An interesting property of this kind of system is that the center of
mass of the system fulfills a closed equation due to the \emph{linear}
character of the force:
\begin{equation}\label{EqCM}
\frac{d^2}{dt^2}\bm{R}(t)=-\omega^2 \bm{R}(t)
\qquad\hbox{with}\qquad \bm{R}(t)\equiv\frac{1}{N}\sum_{i=1}^N\bm{r}_i(t).
\end{equation} 
This
property is independent of the interparticle potential, as it stems from the action-reaction principle and the linear character
of the confining potential. Hence, the center of mass of the system
oscillates around the origin with frequency $\omega $. Of course,
equilibrium can only be reached if $\bm{R}$ and 
$\frac{d}{dt}\bm{R}$ are zero. If this is not the case, the system
can be studied in the non inertial frame of reference in which the center
of mass is at rest. In this frame, the equations of
motion are the same as in the original inertial frame of reference,
again, as a consequence of the 
linear character of the force. This property is specific to the
harmonic potential (isotropic or anisotropic) and it is also valid for
a general interparticle potential. 

\section{Boltzmann equation description of the system}\label{sec3}

In the low density limit, a Boltzmann description of the system is
supposed to be valid. In this case, the state of the system is
described by the one-particle distribution function,
$f(\bm{r},\bm{v},t)$, defined as usual in kinetic theory as
the averaged density of particles in position and velocity space. The
dynamics of the one-particle distribution function are given by
the Boltzmann equation
\begin{equation}\label{BoltzEq}
\left(\frac{\partial}{\partial
    t}+\bm{v}\cdot\frac{\partial}{\partial\bm{r}}
-\omega^2\bm{r}\cdot\frac{\partial}{\partial\bm{v}}\right)
f(\bm{r},\bm{v},t)=J[f|f],   
\end{equation}
where 
\begin{equation}
J[f|f]=\sigma^{d-1}\int d\bm{v}_1\int
d\sig\theta(-\bm{g}\cdot\sig)
\abs{\bm{g}\cdot\sig}(b_{\sig}-1)
f(\bm{r},\bm{v}_1,t)f(\bm{r},\bm{v},t), 
\end{equation}
is the so called collisional contribution. Here, we have introduced the Heaviside step
function, $\theta$, the relative velocity,
$\bm{g}\equiv\bm{v}_1-\bm{v}$, 
and the operator $b_{\sig}$,
that replaces the precollisional velocities into the postcollisional
velocities 
\begin{eqnarray}\label{cra}
\bm{v}'\equiv
  b_{\sig}\bm{v}
&=&\bm{v}+(\bm{g}\cdot\sig)\sig, \\\label{crb}
\bm{v}_1'\equiv
  b_{\sig}\bm{v}_1
&=&\bm{v}_1-(\bm{g}\cdot\sig)\sig. 
\end{eqnarray}

Let us define the averaged values in phase space as 
\begin{equation}
\langle a(\bm{r},\bm{v})\rangle\equiv\frac{1}{N}\int
d\bm{r}\int d\bm{v}a(\bm{r},\bm{v})
f(\bm{r},\bm{v},t). 
\end{equation}
Although $\langle a(\bm{r},\bm{v})\rangle$ depends on time,
this dependence will not be explicitly written. 
By taking moments in the Boltzmann equation, it can be shown that
$\langle r^2\rangle$ and $\langle\bm{r}\cdot\bm{v}\rangle$
satisfy the following first order system of differential equations
\begin{eqnarray}\label{sist1}
\frac{d}{dt}\langle r^2\rangle&=&2
  \langle\bm{r}\cdot\bm{v}\rangle, \\\label{sist2}
\frac{d}{dt}\langle\bm{r}\cdot\bm{v}\rangle&=
&\frac{2e}{m}-2\,\omega^2
  \langle r^2\rangle, 
\end{eqnarray}
where we have introduced the total energy per particle
\begin{equation}
e\equiv\frac{m}{2}\langle v^2\rangle+\frac{m}{2}\omega^2\langle r^2\rangle, 
\end{equation}
that is a constant of the motion. The system given by
Eqs. (\ref{sist1}) and (\ref{sist2}) provides a closed description of 
the moments $\langle r^2\rangle$ and
$\langle\bm{r}\cdot\bm{v}\rangle$, i.e.  they are completely
determined in terms of the initial condition $\langle 
r^2\rangle_0$ and $\langle\bm{r}\cdot\bm{v}\rangle_0$. This,
at first sight, surprising 
feature is a peculiarity of systems confined by an isotropic harmonic
potential and can be intuitively understood as a consequence of the
fact 
that, at the Boltzmann level, 
particles are considered to be point particles
\cite{lbcl2015}. In effect, for one particle Eqs. (\ref{sist1}) and
(\ref{sist2}) trivially hold. Let us then consider a system of two
particles and identify $\langle
r^2\rangle$ with $\frac{1}{2}\left(r_1^2+r_2^2\right)$ and 
$\langle\bm{r}\cdot\bm{v}\rangle$ with 
$\frac{1}{2}\left(\bm{r}_1\cdot\bm{v}_1+\bm{r}_2\cdot\bm{v}_2\right)$. 
Between collisions the result clearly holds. When a collision takes place, the
velocities of the particles change, but $\frac{1}{2}\left(r_1^2+r_2^2\right)$ and 
$\frac{1}{2}\left(\bm{r}_1\cdot\bm{v}_1+\bm{r}_2\cdot\bm{v}_2\right)$ 
do not change if the two particles can be considered to be ``at the
same place'' and momentum is conserved in collisions. Due to this fact
and as energy is conserved in a collision (note that the energy explicitly
appears in Eq. (\ref{sist2})), the system of equations (\ref{sist1})
and (\ref{sist2}) holds for any time. For a 
system composed of an arbitrary number of particles, the argument is
the same as the dynamics can be seen as a sequence of binary
collisions.

The system of differential equations (\ref{sist1}) and (\ref{sist2}) 
is equivalent to the second order differential equation 
\begin{equation}\label{r2EDO}
\frac{d^2}{dt^2}\langle r^2\rangle=\frac{4e}{m}-4\,\omega^2\langle
r^2\rangle, 
\end{equation}
to be solved with the initial condition $\langle r^2\rangle_0$ and 
$\frac{d}{dt}\langle
r^2\rangle_0=2\langle\bm{r}\cdot\bm{v}\rangle_0$. The explicit
solution of Eq. (\ref{r2EDO}) can be written in the form
\begin{equation}
\label{r2Eq}
\langle r^2\rangle=\rho^2+\Delta\cos(2\,\omega t-\varphi), 
\end{equation}
where
\begin{equation}
\rho^2=\frac{e}{m\omega^2}, 
\end{equation}
is the corresponding equilibrium value of $\langle r^2\rangle$, 
\begin{equation}\label{delta}
\Delta=\sqrt{\frac{\langle\bm{r}\cdot\bm{v}\rangle_0^2}{\omega ^2}
+\left(\langle r^2\rangle_0-\frac{e}{m\omega^2}\right)^2}, 
\end{equation}
and $\tan\varphi=\langle\bm{r}\cdot\bm{v}\rangle_0/
\left(\omega\,\langle r^2\rangle_0-\frac{e}{m\,\omega}\right)$. That is, $\langle
r^2\rangle$ oscillates around the equilibrium value, $\rho^2$, 
with angular frequency $2\,\omega$, and with amplitude 
$\Delta$, given by Eq. (\ref{delta}). From this, it is clearly seen
that, in general, equilibrium cannot be reached. It can only be reached
for the exceptional initial conditions in which $\Delta=0$,
i.e. when $\langle\bm{r}\cdot\bm{v}\rangle_0=0$ and $\langle
r^2\rangle$ is equal to the equilibrium value. 

Finally, let us mention that, by taking moments in the Boltzmann
equation, the following relations are obtained
\begin{eqnarray}
\frac{d}{dt}\langle\bm{r}\rangle&=&\langle\bm{v}\rangle, \\
\frac{d}{dt}\langle\bm{v}\rangle&=&-\omega^2\langle\bm{r}\rangle, 
\end{eqnarray}
consistently with the fact that the center of mass of the system obeys
Eq. (\ref{EqCM}). 

\subsection{The breather state}\label{bsSection}
It was shown by Boltzmann that Eq. (\ref{BoltzEq}) admits as a
solution a Gaussian distribution
\begin{equation}\label{fBEq}
f_B(\bm{r},\bm{v},t)=\text{exp}[-\alpha(\bm{r},t)-\beta(\bm{r},t)v^2
-\boldsymbol{\gamma}(\bm{r},t)\cdot\bm{v}], 
\end{equation}
if the coefficients $\alpha$, $\beta$ and $\boldsymbol{\gamma}$ verify
certain conditions \cite{b1909, u63}. We will see that these solutions describe
breathing modes, where a perpetual conversion of kinetic energy and
potential energy operates through a swinglike mechanism. The subscript
$B$ denotes this particular state, the \emph{breather} state. By substituting Eq. (\ref{fBEq})
into the Boltzmann equation, Eq. (\ref{BoltzEq}), it can be shown that $\beta$ has to be
position independent and obeys the following third order differential
equation
\begin{equation}\label{betaEq}
\dddot{\beta}(t)+4\,\omega^2\dot{\beta}(t)=0, 
\end{equation}
where the dot denotes time derivative. The other coefficients must be 
of the form
\begin{eqnarray}
\alpha
  (\bm{r},t)&=&\alpha_0-\dot{\boldsymbol{\gamma}_0}(t)\cdot\bm{r}
+\frac{1}{2}\left[\ddot{\beta}(t)+2\beta(t)\omega^2\right]r^2, \\
\boldsymbol{\gamma}(\bm{r},t)&=&\boldsymbol{\gamma}_0(t)+\bm{J}\times\bm{r}-\dot{\beta}(t)\bm{r}, 
\end{eqnarray}
where $\alpha_0$ and $\bm{J}$ are constants (time and position
independent) and $\boldsymbol{\gamma}_0(t)$ satisfies the second order
differential equation
\begin{equation}
\ddot{\boldsymbol{\gamma}_0}(t)+\omega^2 \boldsymbol{\gamma}_0(t)=\bm{0}. 
\end{equation}
The vectors $\boldsymbol{\gamma}_0(t)$ and $\bm{J}$  describe the position of the
center of mass of the system and the total angular momentum respectively. Taking
into account the remarks of Sec. \ref{secModel}, we can take 
$\boldsymbol{\gamma}_0(t)=\bm{0}$. In addition, we will take
$\bm{J}=\bm{0}$ for simplicity. The analysis for
$\bm{J}\ne\bm{0}$ can be done in a similar fashion. 

It is convenient to introduce the hydrodynamic fields, as usual in
kinetic theory, as the first velocity moments of the distribution
function
\begin{eqnarray}
n(\bm{r},t)&=&\int d\bm{v}f(\bm{r},\bm{v},t), \\
n(\bm{r},t)\bm{u}(\bm{r},t)&=&\int
                                           d\bm{v}\bm{v}f(\bm{r},\bm{v},t),
  \\
\frac{d}{2}n(\bm{r},t)T(\bm{r},t)&=&\int
                                             d\bm{v}\frac{m}{2}
\left[\bm{v}-\bm{u}(\bm{r},t)\right]^2f(\bm{r},\bm{v},t). 
\end{eqnarray} 
These moments are respectively the particle density, momentum density and kinetic energy density. 
In terms of these moments, and under the above mentioned conditions in
which total momentum and total angular momentum are null, the
one-particle distribution of the breather reads
\begin{equation}\label{fb}
f_B(\bm{r},\bm{v},t)=n_B(\bm{r},t)\left[\frac{\beta(t)}{\pi}\right]^{d/2}
\text{exp}\left\{-\beta(t)\left[\bm{v}-\bm{u}_B(\bm{r},t)\right]^2\right\}, 
\end{equation}
with the following hydrodynamic fields 
\begin{eqnarray}
n_B(\bm{r},t)&=&N\left[\frac{a}{4\pi\beta(t)}\right]^{d/2}
\text{exp}\left[-\frac{a}{4\beta(t)}r^2\right],\label{nBEq} \\
\bm{u}_B(\bm{r},t)&=&\frac{\dot{\beta}(t)}{2\beta(t)}\bm{r}.  \label{uBEq}
\end{eqnarray}
Here, $N$ is the total number of particles and the only condition for
$\beta(t)$ is to fulfill Eq. (\ref{betaEq}), whose solution is 
\begin{equation}
\beta(t)=\beta_s+\Delta_{\beta}\cos(2\,\omega t-\varphi). 
\end{equation}
We have also introduced the constant
\begin{equation}\label{defConsA}
a=4\,\omega^2(\beta_s^2-\Delta_{\beta}^2), 
\end{equation}
that is positive by definition. The relation between $\beta(t)$ and
the temperature of the breather, $T_B(t)$, is simply
$T_B(t)=\frac{m}{2\beta(t)}$. 
As the density profile is Gaussian, $\langle
r^2\rangle$ can be calculated in a simple way, obtaining
\begin{equation}\label{r2betaEq}
\langle r^2\rangle=\frac{2d}{a}\beta(t), 
\end{equation}
consistently with Eq. (\ref{r2Eq}) and with the fact that we have
introduced the same notation for the two phases,
$\varphi$. In fact,  Eq. (\ref{r2betaEq}) can be inverted to express
the parameters that define 
$\beta$ ($\beta_s$ and $\Delta_\beta$) in terms of the ones that 
define $\langle r^2\rangle$ ($e$ and $\Delta$), obtaining
\begin{equation}
\beta_s^{-1}=\frac{2(1-q^2)e}{dm}
\end{equation}
with
\begin{equation}
q\equiv\frac{\Delta_{\beta}}{\beta_s}=\frac{m\omega^2\Delta}{e}. 
\end{equation}
We have $0\leq q < 1$ ; $q$ is a meaningful index for quantifying departure from equilibrium, it measures the breather strength: $q=0$ at equilibrium while $q\to 1$ means that $\beta_s \to \infty$, corresponding to a breather of maximal amplitude.
Equivalently, $\beta$ can be explicitly written in terms of $\langle
r^2\rangle$ as 
\begin{equation}\label{betar2Eq}
\frac{d}{2\beta(t)}=\frac{2e}{m}-\omega^2\langle r^2\rangle
-\frac{\langle\bm{r}\cdot\bm{v}\rangle^2}{\langle
                       r^2\rangle}. 
\end{equation}
Nevertheless, it must be stressed 
that, while the time evolution of $\langle r^2\rangle$ is always
described by Eq. (\ref{r2Eq}), the inverse of the temperature, $\beta$,
fulfills Eq. (\ref{betaEq}) only in the breather state. Let us also mention that the breather solution of the Boltzmann equation, $f_B$, although expressed in terms of the hydrodynamic fields, is an \emph{exact} solution of the kinetic equation independently of the values of the fields gradients. In other words, there is no ``hydrodynamic'' approximation behind and the fields of $f_B$ can vary over distances of the order of the mean free path. 

To sum up, the Boltzmann equation admits a time dependent Gaussian
solution completely characterized by $\beta(t)$ or, equivalently, by
$\langle r^2\rangle$. As the dynamics of $\langle r^2\rangle$ are fully
described in terms of the initial conditions
$\langle r^2\rangle_0$, $\langle\bm{r}\cdot\bm{v}\rangle_0$
and $e$, for the considered case in which the center of mass of the
system is at rest and the total angular momentum is zero, the breather
is completely determined in terms of the total number of particles,
$N$, $\langle r^2\rangle_0$, $\langle\bm{r}\cdot\bm{v}\rangle_0$
and $e$. In addition, from the $H$-theorem, it is known that,
independently of the initial condition, $\log
f(\bm{r},\bm{v},t)$ will tend in the long-time limit to be a
collisional invariant \cite{resibois, cercignani}, i.e. 
\begin{equation}\label{colInvB}
\log f(\bm{r},\bm{v},t)\to -\alpha(\bm{r},t)-\beta(\bm{r},t)v^2
-\boldsymbol{\gamma}(\bm{r},t)\cdot\bm{v}.
\end{equation} 
As the breather state is of this form, it turns out that, for
any initial condition, the system will reach in the long-time limit
the only compatible breather with the initial condition, the one
characterized by $N$, 
$\langle r^2\rangle_0$, $\langle\bm{r}\cdot\bm{v}\rangle_0$
and $e$. Hence, in the
context of the studied model, the breather state plays an essential role as it
is the ``attractor'' to which any state will tend to. Moreover,
equilibrium will be reached if and only if the initial condition
fulfills $\langle r^2\rangle_0=\rho^2$ and
$\langle\bm{r}\cdot\bm{v}\rangle_0=0$.

\subsection{Simulation results}\label{simulationsBoltzmann}

In this section we present MD simulations
results of $N$
hard disks (i.e. $d=2$) of mass $m$ and diameter $\sigma$ confined by an harmonic force of constant $k$. In the simulations $m$, $\sigma$ and $k$ are
taken to be unity. When there is a binary
encounter between the 
particles, they collide following the collision rule given by
Eqs. (\ref{cr1}) and (\ref{cr2}). The original event-driven algorithm
for hard spheres \cite{allen} is modified in order to take into account the external force. The
objective is to see if, independently of the initial
condition, the system reaches the breather state studied in the
previous section. Of course, the comparison only makes sense for low
densities where the Boltzmann equation is supposed to accurately
describe the dynamics of the system. In all the simulations performed
in this section, we have taken $N=1000$ 
and the results 
have been averaged over 
$50$ trajectories. The initial condition is generated by placing the particles inside a ring with an inner radius of $R_m$ and an outer radius of $R_M$ as follows: $S$ concentric, equally spaced circles are defined within the ring, with radii ranging from $R_m$ to $R_M$. The same number of particles, $N/S$, is uniformly distributed in each circle. The initial velocity
distribution is Gaussian with null 
total momentum and total angular momentum. In
addition, $\langle\bm{r}\cdot\bm{v}\rangle_0$ is zero and 
$\langle r^2\rangle_0$ is maximum, so that 
$\frac{\langle r^2\rangle_0-\rho^2}{\rho^2}=q$. Let us also define the maximum dimensionless density of a system in equilibrium with energy per particle, $e$, at the Boltzmann level
\begin{equation}\label{phi0Def}
\phi_0=N\left(\frac{d}{2\pi\rho^2}\right)^{d/2}\sigma^d,     
\end{equation}
in terms of which the parameters of the simulations will be expressed. 

\begin{figure}
\begin{center}
\includegraphics[angle=0,width=0.7\linewidth,clip]{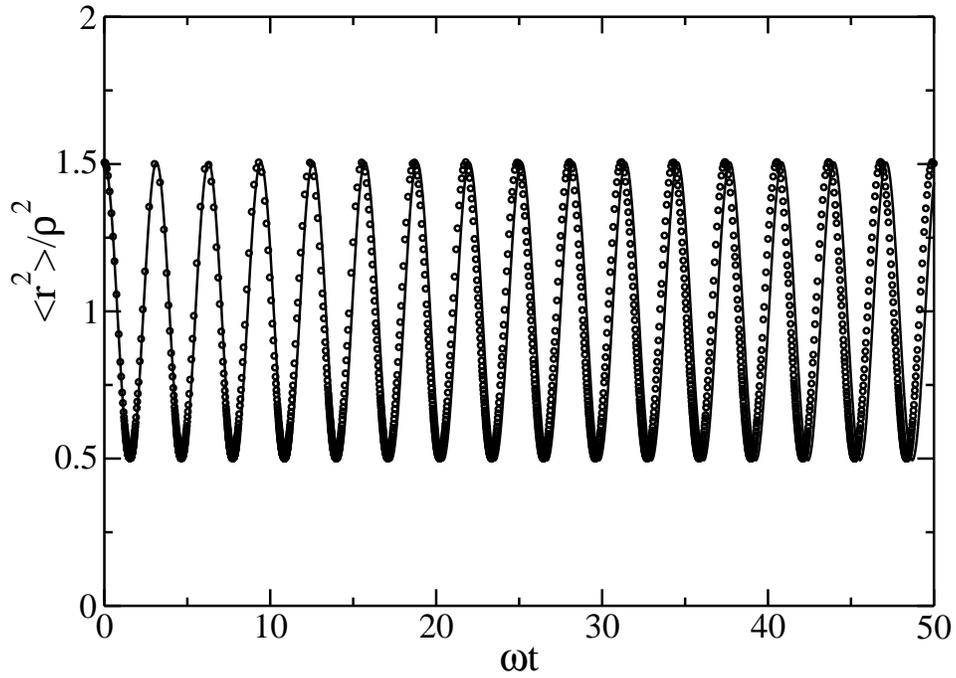}
\end{center}
\caption{$\frac{\langle r^2\rangle}{\rho^2}$ as
a function of the dimensionless time, $\omega t$, for 
$\phi_0=9\times 10^{-3}$ and $q=0.5$. The circles are the Molecular Dynamics simulation results in $d=2$ dimensions and the
solid line the theoretical prediction given by Eq. (\ref{r2Eq}). The number of collisions per particle and per period of oscillation is close to 4. }\label{labelF1}
\end{figure}

\begin{figure}
\begin{center}
\includegraphics[angle=0,width=0.7\linewidth,clip]{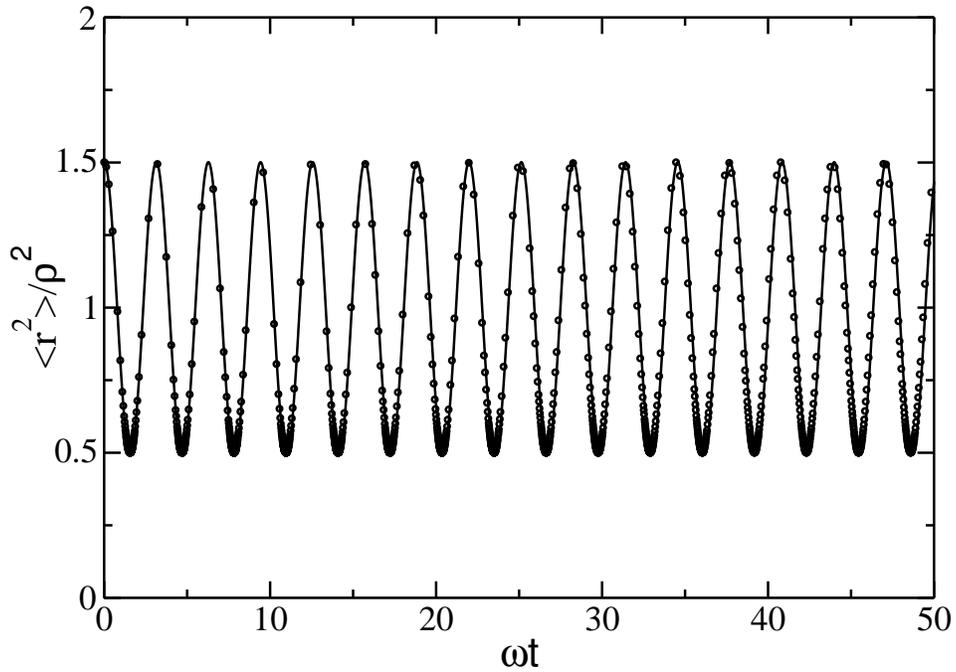}
\end{center}
\caption{Same as in Fig. \ref{labelF1}, but with a smaller density, for $\phi_0=2\times 10^{-3}$. The number of collisions per particle and per period of oscillation is around 3. }\label{labelF1b}
\end{figure}

In Fig. \ref{labelF1},  $\frac{\langle r^2\rangle}{\rho^2}$ is plotted as
a function of the dimensionless time, $\omega t$, for
$\phi_0=9\times 10^{-3}$ and $q=0.5$. 
The circles are the simulation results and the
solid line the theoretical prediction given by Eq. (\ref{r2Eq}). For
the chosen values of the parameters, the density at
the origin (the maximum density) oscillates between
$6\times 10^{-3}\sigma^{-2}$ and $1.8\times 10^{-2}\sigma^{-2}$, so that the system
is supposed to be accurately described by 
the Boltzmann equation. The measurements of $\langle r^2\rangle$ are taken
each $1/30$ collisions per particle. That is why it is seen in the
figure that the density of
circles is higher when $\langle r^2\rangle$ is close to the minimum. 
At those times the temperature and the density are close to the respective maxima and
there are more collisions per particle in a given time window. 
As it is seen, in the time scale of the figure, the
agreement between the simulation results and the theoretical
prediction is very good from the initial time (as expected), 
$\frac{\langle r^2\rangle}{\rho^2}$ oscillating with frequency $2\,\omega$ around
unity. This is so because, as it was 
discussed in Sec. \ref{sec3}, Eq. (\ref{r2Eq}) holds for any initial
condition at any time. Nevertheless, a slight discrepancy between the
frequency seen in the simulations and the theoretical one, $2\,\omega$, is
observed. The measured time average of $\frac{\langle r^2\rangle}{\rho^2}$
is around $1.0036$. It is not plotted because it cannot
be distinguished from the theoretical prediction (unity) in the scale of the
figure. Fig. \ref{labelF1b} shows the same data as Fig. \ref{labelF1}, but for another value of the density, 
$\phi_0=2\times 10^{-3}$. The initial scaled variance is the same
that in the previous case so that $q=0.5$. 
The circles are the simulation results and the
solid line the theoretical prediction given by Eq. (\ref{r2Eq}). As
the energy is larger than in the previous case, the density is smaller
and the Boltzmann prediction is expected to be even better. In fact, in
this case, the discrepancy between the frequency of the simulation
results and the theoretical prediction, $2\,\omega$, cannot be observed in
the scale of the figure. As the measurements of $\langle r^2\rangle$
are also taken
each $1/30$ collisions per particle, the density of points depends on time in a similar way as in Fig. \ref{labelF1}. Let us note that,
for the chosen values of the parameters, the amplitude of the
oscillations is of the order of the mean value ($q=0.5$) and the
dynamics are highly non-linear. As discussed in Sec. \ref{bsSection},
the state of the system is described by an exact solution of the complete
non-linear Boltzmann equation. 

\begin{figure}
\begin{center}
\includegraphics[angle=0,width=0.7\linewidth,clip]{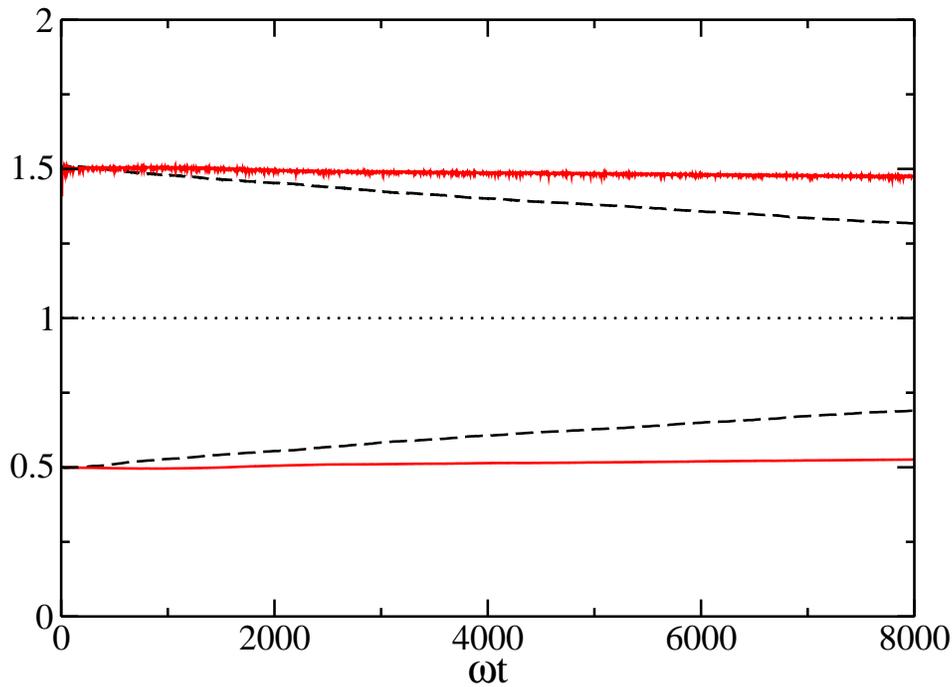}
\end{center}
\caption{Amplitude of oscillations for $\frac{\langle r^2\rangle}{\rho^2}$ as
a function of the dimensionless time (color online). 
The solid line (red) corresponds to
$\phi_0=2\times 10^{-3}$, while the dashed line (black) is for
$\phi_0=9\times 10^{-3}$. The dotted line at unity is a guide to the eye. }\label{labelF2}
\end{figure}

In Fig. \ref{labelF2}, the amplitude of the oscillations of $\frac{\langle
r^2\rangle}{\rho^2}$ is plotted, but over a 
much wider time scale. The solid line (red) corresponds to
$\phi_0=2\times 10^{-3}$, while the dashed line (black) is for
$\phi_0=9\times 10^{-3}$. The point line is at unity and is
plotted only for reference. It is seen that the amplitudes
decay very slowly, with faster decay for the one associated to the larger density. These effects (decay of the amplitude and 
corrections to the frequency and time average value) are
beyond Boltzmann and will be discussed in the following sections. In any
case, there is a wide time scale in which the Boltzmann prediction is
accurate, and that we study in more detail below.

\begin{figure}
\begin{center}
\includegraphics[angle=0,width=0.7\linewidth,clip]{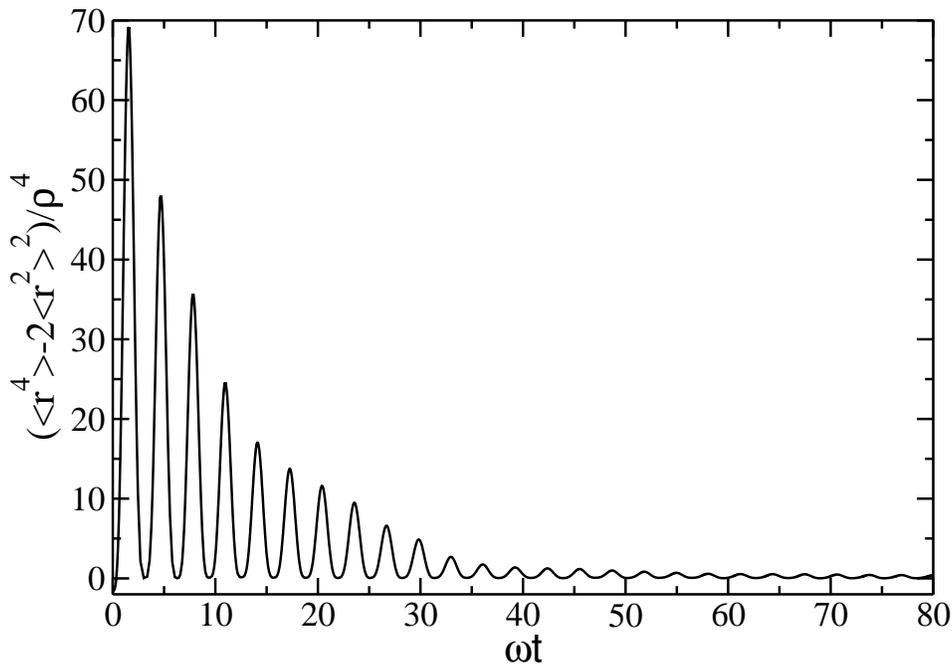}
\end{center}
\caption{MD simulation results for $\frac{\langle r^4\rangle-2\langle r^2\rangle^2}{\rho^4}$ as
a function of the dimensionless time, $\omega t$, for
$\phi_0=9\times 10^{-3}$ and $q=0.5$.   }\label{labelr4}
\end{figure}

\begin{figure}
\begin{center}
\includegraphics[angle=0,width=0.7\linewidth,clip]{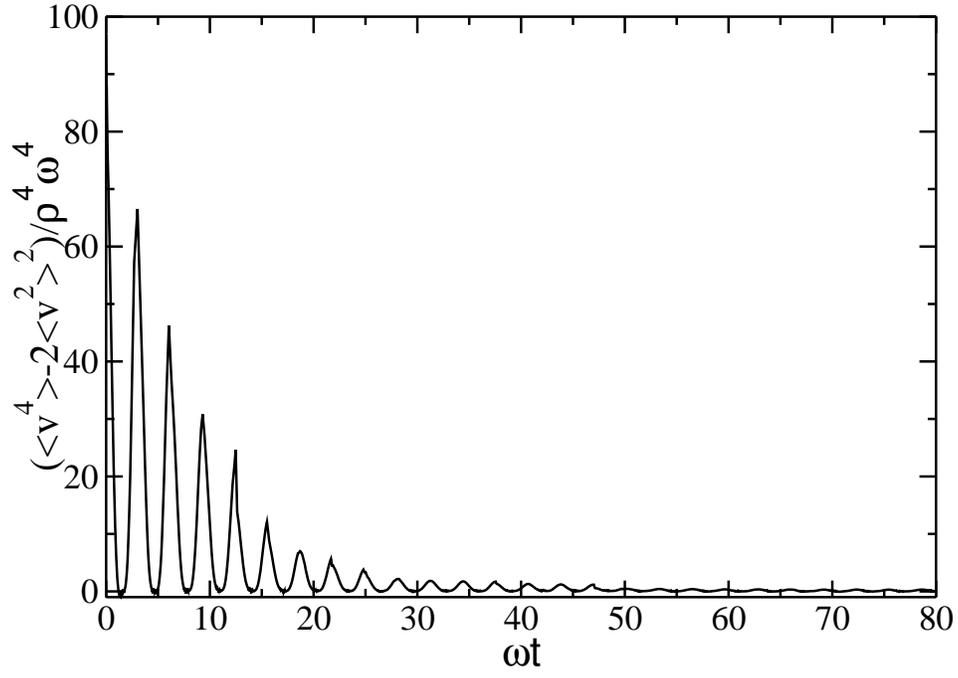}
\end{center}
\caption{MD simulation results for $\frac{\langle v^4\rangle-2\langle v^2\rangle^2}{\omega ^4\rho^4}$ as
a function of the dimensionless time, $\omega t$, for
the same parameters as Fig. \ref{labelr4}.   }\label{labelv4}
\end{figure}

\begin{figure}
\begin{center}
\includegraphics[angle=0,width=0.7\linewidth,clip]{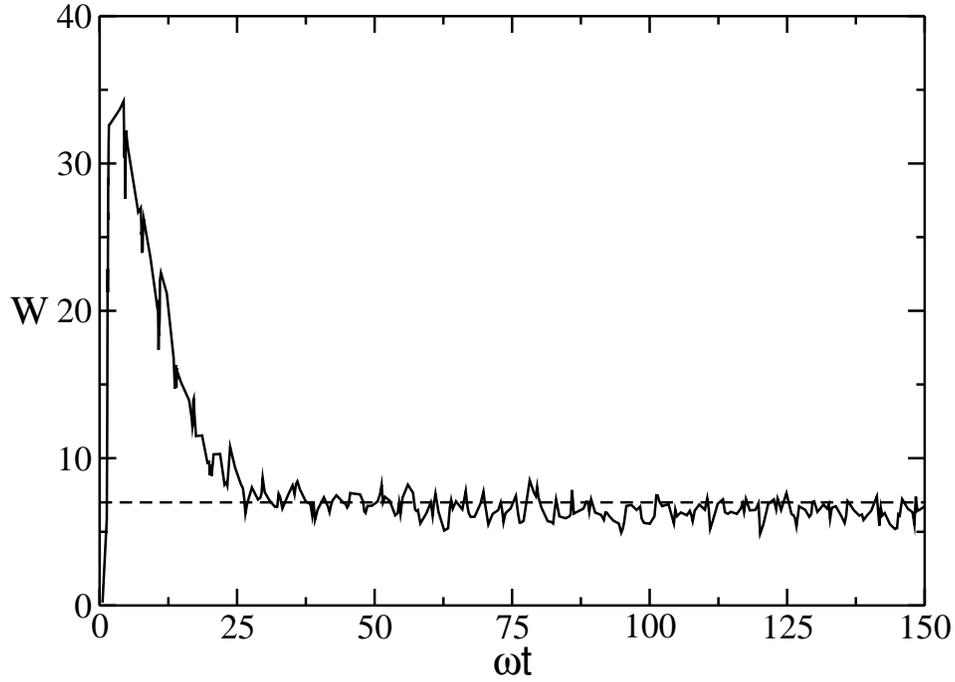}
\end{center}
\caption{$\mathcal{W}$ as a function of the dimensionless time, $\omega t$, for 
the same parameters as Figs. \ref{labelr4} and \ref{labelv4}. The solid line are the simulation results and the dashed
line the theoretical prediction in the breather state. Note that, unlike the fourth moments, $\mathcal{W}$ does not oscillate in the transient to the breather state.}\label{labelFOrigen}
\end{figure}

To see that the system reaches the breather state, we have measured
the fourth position and velocity moments of the distribution function,
$\langle r^4\rangle$ and $\langle v^4\rangle$ respectively. In the
breather state, due to the Gaussian character of the distribution function, these moments are simply related to their respective second moments. In the $d=2$ case, we get
\begin{eqnarray}
&&\langle r^4\rangle_B=2\langle r^2\rangle_B^2, \label{r4Teorica}\\
&&\langle v^4\rangle_B=2\langle v^2\rangle_B^2. \label{v4Teorica}
\end{eqnarray}
In Fig. \ref{labelr4}, MD simulation results for  
$\frac{\langle r^4\rangle-2\langle r^2\rangle^2}{\rho^4}$ are plotted as
a function of the dimensionless time, $\omega t$, for
$\phi_0=9\times 10^{-3}$ and $q=0.5$. It can be seen that, for times $\omega t\sim 50$
(around $100$ collisions per particle), the simulation results are close to zero, indicating that the breather state has been
reached. Similarly, In
Fig. \ref{labelv4}, MD simulation results for  
$\frac{\langle v^4\rangle-2\langle v^2\rangle^2}{\omega ^4\rho^4}$ are plotted as
a function of time, for the same values of the
parameters. Again, for times $\omega t\sim 50$, the simulation results are close to zero, indicating that the breather state has been reached. It is also possible to determine if the system has reached the
breather state by measuring the total number of particles inside a
``small'' sphere in the position and velocity space centered at the origin. By small, it is
understood that the distribution function does not vary appreciably
inside it. This is so because the distribution function of the
breather at the origin is simply $f_B(\bm{0}, \bm{0},
t)=\frac{Na}{4\pi^2}$, that is time-independent (a similar quantity was considered in
\cite{DBB21} to measure the damping of the breathing mode). In
Fig. \ref{labelFOrigen}, the total
number of particles inside a sphere centered at the origin in the
position and velocity space of radius $r_0=50\sigma$ and $v_0=50 \,\omega\,\sigma$
respectively, $\mathcal{W}$, is plotted as a function of the dimensionless time, $\omega t$, for $\phi_0=9\times 10^{-3}$ and $q=0.5$. As the phase-space volume is ``small'', it is
\begin{equation}
\mathcal{W}(t)\approx\Delta\bm{r}\Delta\bm{v}f(\bm{0},\bm{0},t), 
\end{equation}
with $\Delta\bm{r}=\pi r_0^2$ and $\Delta\bm{v}=\pi v_0^2$.  The solid line are the simulation results and the dashed
line the theoretical prediction in the breather state. Although the
quantity fluctuates much more than $\langle r^4\rangle$ and $\langle
v^4\rangle$, it can safely be said that, for $\omega t\sim 25$ (see Fig. \ref{labelFOrigen}), the breather
value has been reached. Note that, unlike the fourth moments, 
$\langle r^4\rangle$ and $\langle v^4\rangle$, $\mathcal{W}$ does not oscillate in the transient to the breather state. 
The results for $\phi_0=2\times 10^{-3}$ and $q=0.5$ are 
similar, finding that, after around $100$ collisions per particle (in
this case $\omega t\sim 120$) $\langle r^2\rangle$, $\langle v^2\rangle$ and
$\mathcal{W}$ reach the corresponding values in the breather state. 

With the aid of the above defined quantities, it is possible to
determine if the breather state has been reached. In particular, the
measured values of the fourth position
and velocity moments indicate that 
the distribution function is Gaussian. The objective now is 
to measure the hydrodynamic fields once the breather state has been reached, in order
to compare them with the theoretical predictions. Concretely, the
spatial domain of the system has been 
divided into rings of internal radius $r$ and external radius 
$r+\Delta r$, that will be denoted as $R(r,\Delta r)$, and three
quantities (functions of $r$) have been measured at different times: (a) the number of
particles in the ring, $\mathcal{M}(r,t)\Delta r$; (b) the averaged radial
component of the velocity in the ring,
$\mathcal{U}(r,t)\equiv\frac{1}{\mathcal{M}(r,t)\Delta r}
\sum_{i\in R(r,\Delta
  r)}\bm{v}_i(t)\cdot\frac{\bm{r}_i(t)}{r_i(t)}$ and; (c) the
  averaged kinetic energy in the ring, 
$\mathcal{E}_k(r,t) \equiv\frac{1}{\mathcal{M}(r,t)\Delta r}
\sum_{i\in R(r,\Delta r)}\frac{m}{2}v_i^2(t)$. The
subindex 
$i\in R(r,\Delta r)$ indicates that the sum is extended over the
  particles that are inside the ring $ R(r,\Delta r)$. From
  Eqs. (\ref{nBEq}) and (\ref{uBEq}), the theoretical expressions for
  the above defined quantities can be calculated in the
  breather. Assuming that $\Delta r$ is small enough
  and for $d=2$, they are: $\mathcal{M}(r,t)=\frac{2N}{\langle
    r^2\rangle}re^{-\frac{r^2}{\langle r^2\rangle}}$, 
$\mathcal{U}(r,t)=\frac{r}{2\langle r^2\rangle}\frac{d \langle
  r^2\rangle}{dt}$ and $\mathcal{E}_k(r,t)=T_B(t)+\frac{m}{8}
\left[\frac{r}{\langle r^2\rangle}\frac{d \langle
  r^2\rangle}{dt}\right]^2$.  We have taken $\Delta r=2.3\sigma$, which 
satisfies the desired condition that the hydrodynamic fields do not vary
appreciably over distances of this order. 

\begin{figure}
\begin{center}
\includegraphics[angle=0,width=0.7\linewidth,clip]{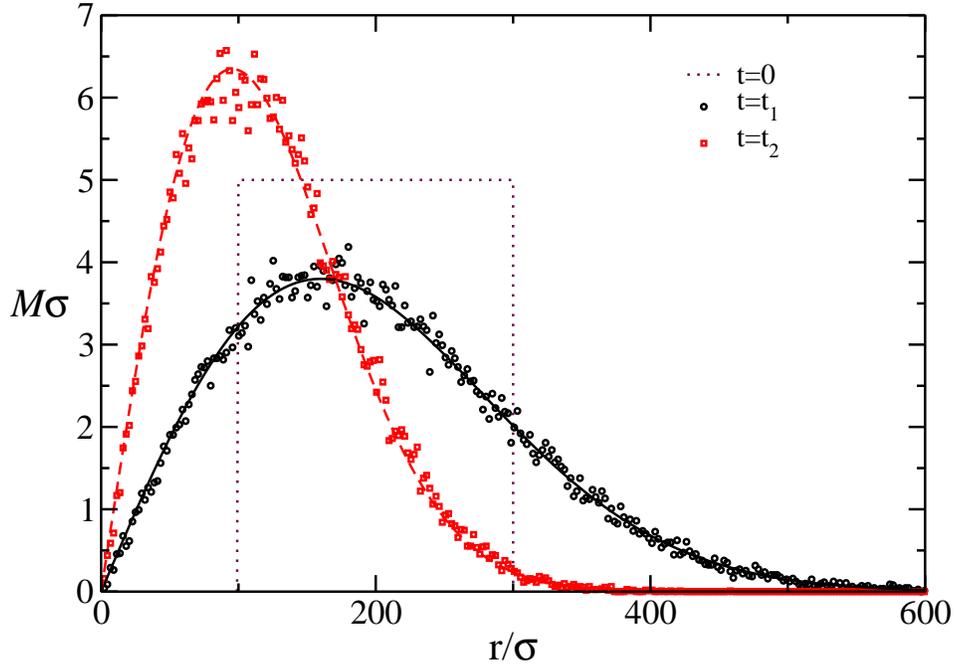}
\end{center}
\caption{$\sigma\mathcal{M}(r,t)$ as a function of $r/\sigma$ for the
  initial time, $t=t_1$ (where 
$\langle r^2\rangle$ is approximately maximum)  and
  $t=t_2$ (where
$\langle r^2\rangle$ is approximately minimum). 
The dots, (black) circles and (red) squares are
the simulation results for the initial condition, $t=t_1$ and
$t=t_2$ respectively. The (black) solid and the (red) dashed lines 
are the theoretical predictions for $t=t_1$ and $t=t_2$
respectively.}\label{labelF3}
\end{figure}

The following MD simulation results are also for
$\phi_0=9\times 10^{-3}$ and $q=0.5$. In Fig. \ref{labelF3},
$\sigma\mathcal{M}(r,t)$ is plotted for the initial 
time, for $\omega t_1=1080$ (where 
$\langle r^2\rangle$ is approximately maximum) and for $\omega t_2$ which
is the value of the available dimensionless time closest to $\omega t_1+\frac{\pi}{2}$ (where 
$\langle r^2\rangle$ is approximately minimum). As mention above, for
these values of the parameters, the system is well in the breather
state for these times. The dots, (black) circles and (red) squares are
the simulation results for the initial condition, $t=t_1$ and
$t=t_2$ respectively. The (black) solid and the (red) dashed lines 
are the theoretical predictions for $t=t_1$ and $t=t_2$
respectively. With the chosen parameters, the values of $R_m$ and
$R_M$ are $100\sigma$ and $300\sigma$ respectively, so that $\mathcal{M}(r,0)$ is
only different from zero for $100\sigma<r<300\sigma$ where it is
constant. It is seen that the agreement between the simulation
results and the theoretical prediction is remarkable. For $t=t_1+\frac{\pi m}{\omega }$ with 
$m\in\mathbb{N}$ and $t$ being not too large for the amplitude of $\langle
r^2\rangle$ to have decayed, a very similar profile to the one
measured at time $t_1$ is obtained. The same occurs for the profiles
at $t=t_2+\frac{\pi m}{\omega }$ and the one measured at time
$t=t_2$. 

\begin{figure}
\begin{center}
\includegraphics[angle=0,width=0.7\linewidth,clip]{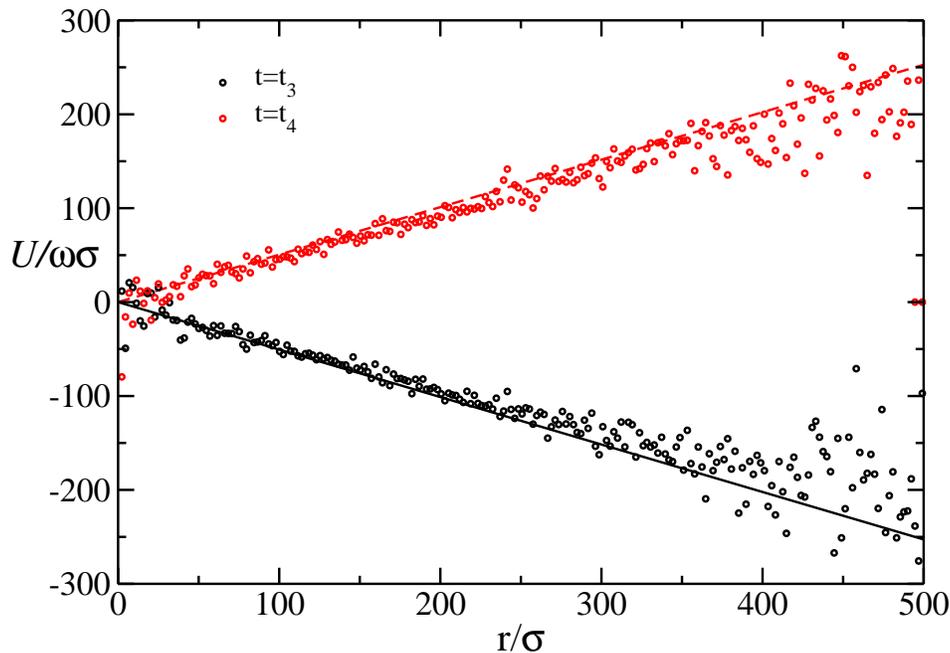}
\end{center}
\caption{$\frac{\mathcal{U}(r,t)}{\sigma \,\omega}$ as a function of
  $r/\sigma $ for
$t=t_3$ and $t=t_4$
where the absolute value of the slope of the velocity profile is
maximum. The (black) circles and (red) squares are the simulation results at time
$t=t_3$ and $t=t_4$ respectively and the (black) solid and (red) dashed lines are
the corresponding theoretical prediction.}\label{labelF4}
\end{figure}

In Fig. \ref{labelF4}, $\frac{\mathcal{U}(r,t)}{\sigma \,\omega}$ is plotted
for the available times closer to 
$t_1+\frac{\pi}{4\,\omega}$ and $t_2+\frac{\pi}{4\,\omega}$ that will be denoted as
$t_3$ and $t_4$ respectively. For these times the absolute value of
the slope of the velocity profile is approximately 
maximum. The (black) circles and (red) squares are the simulation results at time
$t=t_3$ and $t=t_4$ respectively and the (black) solid and (red) dashed lines are
the corresponding theoretical prediction. It is seen that the
agreement between the simulation results and the theoretical
prediction is very good from the origin to $r\sim 400\sigma$. For
$r>400\sigma$ the data becomes more noisy  because there are very few
particles (at those times, the number of
particles that are inside the circle centered at the origin of radius $400\sigma$ is
about $950$ so that the number of particles outside the circle is about
$50$). In fact, as it can be seen in the figure, there are several cells where there are no
particles. $\frac{\mathcal{U}(r,t)}{\sigma \,\omega}$ at times $t=t_1$ and
$t=t_2$ is not shown, but it fluctuates around zero. 

\begin{figure}
\begin{center}
\includegraphics[angle=0,width=0.7\linewidth,clip]{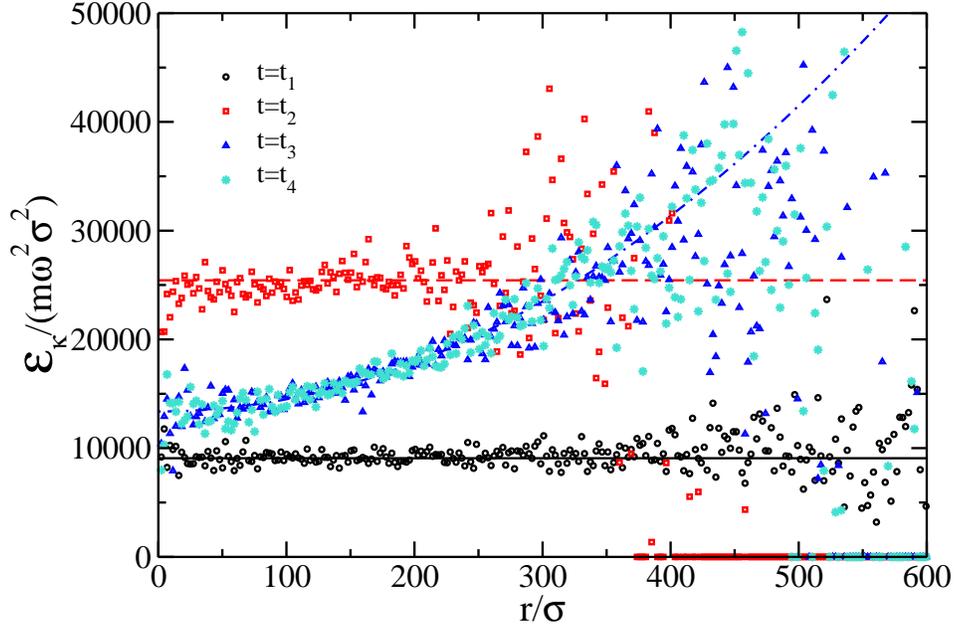}
\end{center}
\caption{$\frac{\mathcal{E}_k(r,t)}{m\omega^2\sigma^2}$ as a function of
  $r/\sigma$ for  $t=t_1$, $t=t_2$, $t=t_3$ and $t=t_4$. The (black) 
circles, (red) squares, (dark blue) triangles and (sky blue) stars are the simulation results for
$t=t_1$, $t=t_2$, $t=t_3$ and $t=t_4$ respectively, and the (black)
solid, (red) dashed and (dark blue) solid-dashed lines the corresponding theoretical
prediction. }\label{labelF5}
\end{figure}

In Fig. \ref{labelF5}, $\frac{\mathcal{E}_k(r,t)}{m\omega^2\sigma^2}$ is
plotted for $t=t_1$, $t=t_2$, $t=t_3$ and $t=t_4$. The 
(black) circles, (red) squares, (dark blue) triangles and (sky blue)
stars are the simulation results for 
$t=t_1$, $t=t_2$, $t=t_3$ and $t=t_4$ respectively, and the (black) 
solid, (red) dashed and (dark blue) solid-dashed lines the corresponding theoretical
prediction. The agreement between the simulation results and the
theoretical prediction is very good from the origin to $r\sim
400\sigma$, $r\sim 200\sigma$ and $r\sim 300\sigma$ for the times
$t=t_1$, $t=t_2$ and $t=t_3$ (the same for $t=t_4$) respectively. As
in the previous case of $\mathcal U$, above these values of the
distance to the origin, the data
become very noisy. 

All these results indicate that for times of the order of 
$t_1$ the system is already in the breather state where it remains for
long time. Similar results are obtained starting with different
initial conditions and for different values of the parameters if the
density is small enough, proving the attractive character of the
breather state at the Boltzmann level.

\section{Breathers beyond the Boltzmann equation description}\label{sec4}

The objective of this section is the study of the existence of the
breather states beyond the low density limit. The question we want to
address is: for the considered model at finite densities, is a breather
state or, on the contrary, equilibrium reached in the long-time limit? 

We have shown in Sec. \ref{sec3} that one of the essential features not to reach
equilibrium in the low density limit is the fact that we have a closed
equation for $\langle r^2\rangle$, Eq. (\ref{r2EDO}). The solution of
this equation 
is given by Eq. (\ref{r2Eq}), where it is manifestly seen that
equilibrium can only be reached for a particular class of initial
conditions. It was also shown in Sec. \ref{sec3} by intuitive
arguments that, to have the above mentioned closed description, it is
essential that the particles be point-like. Hence, we expect not
to have a closed description for $\langle r^2\rangle$ in a finite
density system. On the other hand, for a finite density system of hard
spheres, the so-called Enskog equation is supposed to describe the dynamics of
the system if the density is ``moderate'' \cite{e22, vbe73, resibois}. The structure of the Enskog
equation is similar to that of the Boltzmann equation, but with the collisional
term modified in such a way that the size of the particles as well
as the existence of spatial correlations between the colliding
particles are considered. For this equation, a $H$-theorem has been
proved \cite{r78, mgb18} but, in this case, the condition for the
long-time behavior of $\log f$ is 
\begin{equation}\label{colInvE}
\log f(\bm{r},\bm{v},t)\to -\alpha(\bm{r},t)-\beta(t)v^2
-\boldsymbol{\gamma}(t)\cdot\bm{v}
-\boldsymbol{\delta}(t)\cdot\bm{r}\times\bm{v}, 
\end{equation}
where, in contrast to Eq. (\ref{colInvB}), $\beta$ and
$\boldsymbol{\gamma}$ 
depend only on time but not on the position. The new function,
$\boldsymbol{\delta}$, is included here because total angular momentum
is conserved in a collision (in the Boltzmann framework this term is
included in the $\boldsymbol{\gamma}$ that can depend on
$\bm{r}$). The difference between Eq. (\ref{colInvB}) and
(\ref{colInvE}) is due to the fact that, at the Enskog level, 
$f(\bm{r}+\boldsymbol{\sigma}, \bm{v})$ is not approximated by 
$f(\bm{r}, \bm{v})$ (as it is done at the Boltzmann level in
which the size of the particles is neglected). Then, in a collision, it
is taken into account that particles are located at different places. For 
example, $\beta(\bm{r},t)\bm{v}_1
+\beta(\bm{r}+\boldsymbol{\sigma},t)\bm{v}_2\ne 
\beta(\bm{r},t)\bm{v}_1'
+\beta(\bm{r}+\boldsymbol{\sigma},t)\bm{v}_2'$, while  
$\beta(t)\bm{v}_1
+\beta(t)\bm{v}_2= 
\beta(t)\bm{v}_1'
+\beta(t)\bm{v}_2'$ due to conservation of total
momentum in a collision (the same occurs with $\boldsymbol{\gamma}$ and
$\boldsymbol{\delta}$). The fact that, at the Enskog level,
$\boldsymbol{\gamma}$ can only depend on time forbids the existence of
the breather and equilibrium is always reached. 

For the above mentioned reasons, it is expected that the system will always
reach equilibrium in the long-time limit. Nevertheless, we would also
expect that, if the system is dilute enough, the Boltzmann description
will provide a good approximation of the dynamics and
the system will reach a state close to the breather where it will
remain for some time. In order to study this situation, we will start
by a Liouville description of the system. The first equation of the
BBGKY hierarchy is \cite{cercignani}
\begin{equation}\label{Bbgky1}
\left(\frac{\partial}{\partial
    t}+\bm{v}\cdot\frac{\partial}{\partial\bm{r}}
-\omega^2\bm{r}\cdot\frac{\partial}{\partial\bm{v}}\right)
f(\bm{r},\bm{v},t)=L[f_2],   
\end{equation}
where 
\begin{equation}
L[f_2]=\sigma^{d-1}\int d\bm{v}_1\int d\sig
\abs{\bm{g}\cdot\sig} \left[\theta(\bm{g}\cdot\sig)b_{\sig}
-\theta(-\bm{g}\cdot\sig)\right]
f_2(\bm{r}+\boldsymbol{\sigma},\bm{v}_1,\bm{r},\bm{v},t), 
\end{equation}
being
$f_2(\bm{r}+\boldsymbol{\sigma},\bm{v}_1,\bm{r},\bm{v},t)$ 
the two-particle distribution function of the colliding
particles. Note that the Boltzmann equation is obtained by making the
approximation 
$f_2(\bm{r}+\boldsymbol{\sigma},\bm{v}_1,\bm{r},\bm{v},t)\approx
f(\bm{r},\bm{v}_1,t) f(\bm{r},\bm{v},t)$,
i.e. neglecting position and velocity correlations and the variation
of the one-particle distribution function in distances of the order of
the diameter of the particles. 

Performing a similar analysis to the one made in the previous section
with the Boltzmann equation, i.e. by taking moments in Eq. (\ref{Bbgky1}),
one obtains (see Appendix \ref{app1})
\begin{eqnarray}\label{sistM1}
\frac{d}{dt}\langle r^2\rangle&=&2
  \langle\bm{r}\cdot\bm{v}\rangle, \\\label{sistM2}
\frac{d}{dt}\langle\bm{r}\cdot\bm{v}\rangle&=&\frac{2e}{m}-2\,\omega^2
  \langle r^2\rangle+\frac{1}{mN}\int d\bm{r}\text{Tr}\mathcal{P}^{(c)}(\bm{r},t), 
\end{eqnarray}
where $\mathcal{P}^{(c)}(\bm{r},t)$ is the collisional
contribution to the pressure tensor. Its explicit expression is
\begin{equation}\label{colPreTen}
\mathcal{P}_{ij}^{(c)}(\bm{r},t)=\frac{m}{2}\sigma^d\int
d\bm{v}_1\int d\bm{v}\int_0^1 d\lambda\int d\sig
\theta(-\bm{g}\cdot\sig)f_2\left[\bm{r}_1(\lambda, \sig),
\bm{v}_1, \bm{r}_2(\lambda, \sig), \bm{v}, t\right]
(\bm{g}\cdot\sig)^2\widehat{\sigma}_i\widehat{\sigma}_j, 
\end{equation}
where
\begin{equation}
\bm{r}_1(\lambda, \sig)=\bm{r}+\lambda\boldsymbol{\sigma},
\quad
\bm{r}_2(\lambda,
\sig)=\bm{r}+(\lambda-1)\boldsymbol{\sigma}. 
\end{equation}
Physically, $\mathcal{P}^{(c)}(\bm{r},t)$ represents the
contribution to the flux of momentum due to collisions
\cite{mgb18}. In a hard sphere system, there is flux of momentum
through a given surface due to particles that cross the surface, 
$\mathcal{P}^{(k)}(\bm{r},t)$, and due to collisions between
particles without crossing the surface (the two particles are in
opposite sites of the surface, do not cross the surface, but
interchange momentum in a collision). The first contribution is
described simply by the tensor $\mathcal{P}_{ij}^{(k)}(\bm{r},t)=
m\int d\bm{v}[v_i-u_i(\bm{r},t)][v_i-u_i(\bm{r},t)]
f(\bm{r},\bm{v},t)$, while the second,
$\mathcal{P}_{ij}^{(c)}(\bm{r},t)$, is given by Eq. (\ref{colPreTen}). 

Eqs. (\ref{sistM1}) and (\ref{sistM2}) are exact but, in contrast to
the Boltzmann ones given by (\ref{sist1}) and (\ref{sist2}), they
are not closed. They depend on 
$\int d\bm{r}\text{Tr}\mathcal{P}^{(c)}(\bm{r},t)$ that is
unknown. Moreover, this new term is a \emph{collisional contribution}
which arises from the fact that, in a collision, particles are not at
the same place due to their finite size (see Eq. (\ref{colPreTen})). So,
everything is consistent with the
intuitive explanation of Eq. (\ref{r2EDO}) in which it was essential
that the particles were ``at the same place'' in a collision in order
to have a closed equation for $\langle r^2\rangle$: for finite size
particles the evolution equation for $\langle r^2\rangle$ is not
closed due to a collisional contribution that vanishes for point
particles. In fact, an intuitive derivation of Eqs. (\ref{sistM1}) and
(\ref{sistM2}) can be made in the same lines as in the Boltzmann
context. In effect, in this case
$\sum_{i=1}^N\bm{r}_i\cdot\bm{v}_i$ changes 
instantaneously in a collision. Concretely, 
$\bm{r}_i\cdot\bm{v}_i'+\bm{r}_j\cdot\bm{v}_j'
-\bm{r}_i\cdot\bm{v}_i-\bm{r}_j\cdot\bm{v}_j=
-\sigma\bm{v}_{ji}\cdot\sig$, where $\sig$ has been defined taking
the particle $i$ as the tagged particle. Note that the increment is always
positive because $\sigma\bm{v}_{ji}\cdot\sig<0$ when a collision
is going to take place. Hence, the
change of $\sum_{i=1}^N\bm{r}_i\cdot\bm{v}_i$ in a time window $\Delta t$ due
to collisions is the sum of  $-\sigma\bm{v}_{ji}\cdot\sig$ for all
the collisions taking place in $\Delta t$. The average of this quantity
is precisely the new contribution 
$\frac{\Delta t}{mN}\int d\bm{r}\text{Tr}\mathcal{P}^{(c)}(\bm{r},t)=
\frac{\Delta t\sigma^d}{2N}\int d\bm{r}\int d\bm{v}_1\int
d\bm{v}_2\int
d\sig\theta(-\bm{v}_{12}\cdot\sig)(\bm{v}_{12}\cdot\sig)^2
f_2(\bm{r}+\boldsymbol{\sigma}, \bm{v}_1, \bm{r},
\bm{v}_2,t)$. 

In order to study Eqs. (\ref{sistM1}) and (\ref{sistM2}) some
approximations have to be done. Firstly, we will assume molecular
chaos, i.e.
\begin{equation}\label{mcApp}
f_2(\bm{r}+\boldsymbol{\sigma}, \bm{v}_1,
\bm{r},\bm{v},t)\approx
f(\bm{r}+\boldsymbol{\sigma}, \bm{v}_1,t)
f(\bm{r}, \bm{v},t), 
\end{equation}
without the approximation 
$f(\bm{r}+\boldsymbol{\sigma}, \bm{v}_1,t)\approx
f(\bm{r}, \bm{v}_1,t)$. This is the simplest approximation
that still takes into account the fact that, in a collision, particles
are not at the same place. To continue, we need to express the
distribution function as functionals of $\langle r^2\rangle$ and 
$\langle\bm{r}\cdot\bm{v}\rangle$. The simulation results of
the previous section show that, for the considered values of the
parameters, the one particle distribution function
of the finite density system is very well characterized by the
distribution function of the breather at the Boltzmann level. This is
so on the time scale in which the amplitude of the breather does not
vary. For this reason, we assume that the distribution function has
the same functional form in terms of $\langle r^2\rangle$ and 
$\langle\bm{r}\cdot\bm{v}\rangle$ as the distribution
function of the breather at the Boltzmann level. Taking into account
Eqs. (\ref{fb}), (\ref{nBEq}) and (\ref{uBEq}), the distribution is
then approximated by
\begin{equation}\label{dfApp}
f(\bm{r},\bm{v},t)\approx n(\bm{r},t)\left[\frac{\beta(t)}{\pi}\right]^{d/2}
e^{-\beta(t)\left[\bm{v}-\bm{u}(\bm{r},t)\right]^2}, 
\end{equation}
with
\begin{eqnarray}
n(\bm{r},t)&=&N\left[\frac{d}{2\pi\langle
    r^2\rangle}\right]^{d/2}e^{-\frac{d}{2\langle r^2\rangle}r^2}, \label{nEq}\\
\bm{u}(\bm{r},t)&=&
\frac{\langle\bm{r}\cdot\bm{v}\rangle}{\langle
                            r^2\rangle}\bm{r}.  \label{uEq}
\end{eqnarray}
The inverse of the temperature, $\beta$, is given in terms of  $\langle r^2\rangle$ and 
$\langle\bm{r}\cdot\bm{v}\rangle$ by Eq. (\ref{betar2Eq})
\begin{equation}
\frac{d}{2\beta(t)}=\frac{2e}{m}-\omega^2\langle r^2\rangle
-\frac{\langle\bm{r}\cdot\bm{v}\rangle^2}{\langle
                       r^2\rangle}. \label{bEq}
\end{equation}
As the distribution function of the breather is an exact solution of the Boltzmann equation in all the regimes, i.e. from the collisionless to the hydrodynamic regime, our approximation is expected to be valid in all the regimes aswell. 

Taking into account molecular chaos and inserting Eq. (\ref{dfApp})
into Eq. (\ref{colPreTen}), the expression for  $\text{Tr}\mathcal{P}^{(c)}$
to first order in the gradients is (see Appendix \ref{app2})
\begin{equation}\label{TrPc}
\text{Tr}\mathcal{P}^{(c)}(\bm{r},t)\approx
\text{Tr}\mathcal{P}^{(c,0)}(\bm{r},t)+
\text{Tr}\mathcal{P}^{(c,1)}(\bm{r},t), 
\end{equation}
where the zeroth and first order terms are
\begin{equation}\label{TrPc0}
\text{Tr}\mathcal{P}^{(c,0)}(\bm{r},t)=\frac{\pi^{d/2}}{\Gamma\left(\frac{d}{2}\right)}
n^2(\bm{r},t)\sigma^dT(t), 
\end{equation}
and
\begin{equation}\label{TrPc1}
\text{Tr}\mathcal{P}^{(c,1)}(\bm{r},t)=
-\frac{2\pi^{\frac{d-1}{2}}}{d\Gamma\left(\frac{d}{2}\right)}
n^2(\bm{r},t)\sigma^{d+1}\left[mT(t)\right]^{1/2}
\nabla\cdot\bm{u}(\bm{r},t), 
\end{equation}
respectively. Note that $\text{Tr}\mathcal{P}^{(c)}$ is proportional
to $n^2(\bm{r},t)$ consistently with the fact that we are beyond
the Boltzmann framework. 

Let us remark that the same result is
obtained in a simpler way (without any calculation) if, for
$\mathcal{P}^{(c)}$, we take the hydrodynamics expression at the 
Enskog level to first order in the gradients and neglect position
correlation \cite{ferziger, gbs18} (taking the pair
correlation function equal to unity) as it was done in \cite{gmgt24}. In effect, the hydrodynamic 
expression for $\mathcal{P}^{(c)}$ to first order in the gradients is \cite{resibois} 
\begin{equation}
\mathcal{P}^{(c)}_{ij}=nT\Delta
p^*\delta_{ij}-\eta_c\left(\frac{\partial u_j}{\partial x_i}
+\frac{\partial u_i}{\partial x_j}-\frac{2}{d}\nabla\cdot\bm{u}\delta_{ij}\right)
-\nu \nabla\cdot\bm{u}\delta_{ij}, 
\end{equation}
where $\Delta p^*$ is the excess pressure with respect to the ideal
one, $nT$, $\eta_c$ is the collisional contribution to the shear
viscosity, $\nu$ is the bulk viscosity and $x_i$ denotes the $i$-th
component of $\bm{r}$. The explicit expression for $\Delta p^*$
and $\nu$ are (the coefficient $\eta_c$ is not needed in the
subsequent analysis)
\begin{equation}
\Delta p^*=\frac{\pi^{d/2}}{d\Gamma\left(\frac{d}{2}\right)}g_2n\sigma^d, 
\end{equation}
and 
\begin{equation}\label{bulkviscosityhs}
\nu=\frac{2\pi^{\frac{d-1}{2}}}{d^2\Gamma\left(\frac{d}{2}\right)}
g_2n^2\sigma^{d+1}(mT)^{1/2}, 
\end{equation}
where $g_2$ is the pair correlation function at contact. 
The trace of the tensor is 
\begin{equation}\label{TrPcE}
\text{Tr}\mathcal{P}^{(c)}=dnT\Delta p^*-d\nu\nabla\cdot\bm{u}. 
\end{equation}
Taking into account the explicit expressions for $\Delta p^*$ and
$\nu$ given above and approximating the two-pair
correlation function by unity, Eq. (\ref{TrPcE}) reduces to
Eq. (\ref{TrPc}) with the zeroth order and first order in the
gradients contribution given by Eqs. (\ref{TrPc0}) and (\ref{TrPc1})
respectively. The coincidence between the kinetic
theory approach and the hydrodynamic approach deserves some comments as it is not
incidental: in the kinetic theory approach the distribution function
is Gaussian, while in the hydrodynamic one the distribution 
function to first order in the gradients is not
Gaussian. Nevertheless, the non
gaussianities contribution to $\mathcal{P}^{(c)}$ are all immersed in
the shear viscosity, that does not appear in the trace. Let us also remark that, in the hydrodynamic approach, it is implicitly assumed that the fields vary over distances much larger than the mean free path while, in the kinetic approach, it is only assumed that the fields vary smoothly over distances of the order of the diameter of the particles (so that the expansion given by Eq. (\ref{TrPc}) makes sense).  

Finally, taking into account the expressions for the
hydrodynamic fields, Eqs. (\ref{nEq}), (\ref{uEq}) and (\ref{bEq}),
the spatial integral of the pressure tensor can be evaluated,
obtaining 
\begin{eqnarray}
\int d\bm{r}\text{Tr}\mathcal{P}^{(c,0)}(\bm{r},t)&=&
\frac{\pi^{d/2}}{2^{\frac{d}{2}}\Gamma\left(\frac{d}{2}\right)}
NT(t)\phi[\langle r^2\rangle], \label{ptc0}\\
\int d\bm{r}\text{Tr}\mathcal{P}^{(c,1)}(\bm{r},t)&=&
-\frac{\pi^{\frac{d-1}{2}}}{2^{\frac{d-2}{2}}\Gamma\left(\frac{d}{2}\right)}N\sigma
\sqrt{mT(t)}\frac{\langle\bm{r}\cdot\bm{v}\rangle}{\langle
                                                              r^2\rangle}
\phi[\langle  r^2\rangle], \label{ptc1}
\end{eqnarray}
where the value of the maximum density (the one at the origin) at the
Boltzmann level at time $t$ has been introduced
\begin{equation}
\phi[\langle  r^2\rangle]=N\left[\frac{d}{2\pi\langle
    r^2\rangle}\right]^{d/2}\sigma^d. 
\end{equation}
Eqs. (\ref{sistM1}) and (\ref{sistM2}) with 
$\int d\bm{r}\text{Tr}\mathcal{P}^{(c)}(\bm{r},t)$ given by
Eqs. (\ref{ptc0}) and (\ref{ptc1}) form a closed set of first order
non-linear differential equations for $\langle r^2\rangle$ and
$\langle\bm{r}\cdot\bm{v}\rangle$. They are valid arbitrary far from
equilibrium if the hypothesis we have postulated are valid, i.e. if 
Eqs. (\ref{mcApp}), (\ref{dfApp}) and (\ref{TrPc}) hold. In the next section, we will analyze the
equation for states close to equilibrium. 

\subsection{Close to equilibrium states}\label{cEqTheory}

In this section, the dynamics of the system are studied for states
close to thermal equilibrium. It is convenient to work with the
second order differential equation for $\langle r^2\rangle$ equivalent
to the system of Eqs. (\ref{sistM1}) and
(\ref{sistM2}) that reads
\begin{equation}\label{r2EqEnskog}
\frac{d^2}{dt^2}\langle r^2\rangle=\frac{4e}{m}-4\,\omega^2
  \langle r^2\rangle+\frac{2}{mN}\int d\bm{r}\text{Tr}\mathcal{P}^{(c)}(\bm{r},t), 
\end{equation}
where $\int d\bm{r}\text{Tr}\mathcal{P}^{(c)}(\bm{r},t)$ is
given by Eqs. (\ref{ptc0}) and (\ref{ptc1}) with
$\langle\bm{r}\cdot\bm{v}\rangle=\frac{1}{2}\frac{d}{dt}\langle
r^2\rangle$. Eq. (\ref{r2EqEnskog}) admits a stationary solution (the equilibrium
solution), $\langle r^2\rangle_e$, that satisfies 
\begin{equation}
\frac{2e}{m}-2\,\omega^2 \langle r^2\rangle_e
+\frac{\pi^{\frac{d}{2}}}{2^{\frac{d}{2}}\Gamma\left(\frac{d}{2}\right)}
\phi[\langle r^2\rangle_e]\frac{T_e}{m}=0, 
\end{equation}
with $T_e=\frac{2}{d}e-\frac{m\omega^2}{d}\langle r^2\rangle_e$. $\langle
r^2\rangle_e$ differs from the corresponding Boltzmann value,
$\rho^2=\frac{e}{m\omega^2}$, due to excluded volume effects. Assuming
that the density is small, $\langle r^2\rangle_e$ should be close to
$\rho^2$. It is convenient to introduce the deviation
\begin{equation}
\delta \langle r^2\rangle_e\equiv\langle r^2\rangle_e-\rho^2, 
\end{equation} 
that, to linear order, can be calculated, obtaining
\begin{equation}\label{deltar2}
\delta \langle
r^2\rangle_e=\frac{\pi^{\frac{d}{2}}}{d2^{\frac{d}{2}+1}
\Gamma\left(\frac{d}{2}\right)}\phi_0\rho^2, 
\end{equation}
where $\phi_0$ is the value of the maximum density of a system in
equilibrium with energy per particle, $e$, at the Boltzmann level, defined in Eq. (\ref{phi0Def}). Of course, 
$\phi_0=\phi[\rho^2]$. 
Remarkably, the prediction for $\langle r^2\rangle_e$ given by
Eq. (\ref{deltar2}) coincides with the equilibrium one in the first
virial approximation (see Appendix \ref{app3}). 

To study the dynamics, we introduce the deviation of $\langle
r^2\rangle$ around the actual equilibrium value
\begin{equation}
x\equiv\langle r^2\rangle-\langle r^2\rangle_e. 
\end{equation}
To linear order in $x$ and for low densities, we have
\begin{eqnarray}\label{TrPc0lx}
\frac{2}{mN}\int
  d\bm{r}\text{Tr}\mathcal{P}^{(c,0)}(\bm{r},t)&\approx&
\frac{2\pi^{d/2}}{2^{d/2}\Gamma\left(\frac{d}{2}\right)}
\phi[\langle r^2\rangle_e]\frac{T_e}{m}
-\frac{(d+2)\pi^{d/2}}{d2^{d/2}\Gamma\left(\frac{d}{2}\right)}\phi_0\omega^2x,
  \\\label{TrPc1lx}
\frac{2}{mN}\int
  d\bm{r}\text{Tr}\mathcal{P}^{(c,1)}(\bm{r},t)&\approx&
-\frac{2\pi^{\frac{d-1}{2}}}{d^{1/2}2^{d/2}\Gamma\left(\frac{d}{2}\right)}\phi_0\frac{\sigma}{\rho}
  \omega\dot{x}.  
\end{eqnarray}
Taking into account the above relations, by linearizing
Eq. (\ref{r2EqEnskog}) around the equilibrium solution, 
one finds that, for low densities, $x$ satisfies the following second order linear
differential equation with constant coefficients
\begin{equation}\label{edox}
\ddot{x}+\frac{2}{\tau}\dot{x}+\Omega ^2x=0, 
\end{equation}
with 
\begin{equation}\label{eqTau}
\omega\tau=\frac{d^{1/2}2^{d/2}\Gamma\left(\frac{d}{2}\right)}{\pi^{\frac{d-1}{2}}}
\frac{\rho}{\phi_0\sigma}, 
\end{equation}
and
\begin{equation}\label{eqW}
\left(\frac{\Omega}{\omega}\right)^2=4+\frac{(d+2)\pi^{d/2}}{d2^{d/2}\Gamma\left(\frac{d}{2}\right)}\phi_0. 
\end{equation}
The solution of the differential equation can be written in terms of
the roots of the characteristic equation, $m^2+\frac{2}{\tau}m+\Omega  ^2=0$, 
\begin{equation}
m=-\frac{1}{\tau}\pm
i\sqrt{\Omega  ^2-\left(\frac{1}{\tau}\right)^2}\approx
-\frac{1}{\tau}\pm i\Omega  , 
\end{equation}
where, again, the linear approximation in $\phi_0$ has been
made. Then, within this approximation, the solution of
Eq. (\ref{edox}) can be written as
\begin{equation}\label{eqFundame}
x(t)=x(0)e^{-\frac{t}{\tau}}\cos(\Omega   t-\varphi), 
\end{equation}
with the phase, $\varphi$, depending on the initial condition. 
The coefficient $\tau$, given by Eq. (\ref{eqTau}), is identified as the relaxation time of the amplitude of the oscillations, and $\Omega$, given by Eq.(\ref{eqW}), is the frequency of the oscillations. This is the main result of the paper as it
lets us understand the origin of the relaxation to equilibrium as a
consequence of density corrections with respect to the Boltzmann
framework. The relaxation time, $\tau$, depends on the two parameters that define the equilibrium state, $\phi_0$ and $\rho/\sigma$ (or, equivalently, $N$ and $e$). In the conditions we are working (for low densities), the relaxation is
very slow, as the relaxation time goes as $\tau\propto
1/\phi_0$. The divergence in the low density limit is 
consistent with the Boltzmann description, in which equilibrium is never reached. On the other hand, the frequency of the oscillations,
$\Omega  $, is ``renormalized'' with respect to the Boltzmann prediction,
$2\,\omega$, being the former one always larger than the last one. 
The origin of the
renormalization of the frequency and of the damping can be identified: from Eq. (\ref{TrPc0lx}), 
it is seen that the renormalization of the frequency (and also of the time average of the oscillations) comes from 
$\text{Tr}\mathcal{P}^{(c,0)}$ while, from Eq. (\ref{TrPc1lx}), it is seen that the damping comes
from $\text{Tr}\mathcal{P}^{(c,1)}$. These results are equivalent to the ones obtained in \cite{gmgt24} by hydrodynamic arguments where $\text{Tr}\mathcal{P}^{(c,0)}$ is associated to the excess pressure and $\text{Tr}\mathcal{P}^{(c,1)}$ to the bulk viscosity. Concretely, in \cite{gmgt24} the relaxation time, $\tau$, was identified as a functional of the bulk viscosity in equilibrium, $\nu_e(\mathbf{r})$, in the form
\begin{equation}
\tau=\frac{2mN\rho^2}{d^2\int d\mathbf{r}\nu_e(\mathbf{r})}, 
\end{equation}
that coincides with Eq. (\ref{eqTau}). Eq. (\ref{eqFundame}) highlights the two time scales of the state of
the system: the one related to the oscillations, $\omega ^{-1}$, that is fast and it is given by the external force, and $\tau$ that controls the decay of the amplitude of the oscillations and that is much slower than the previous one.

\subsection{Simulation results}

The objective of this section is to compare the theoretical
predictions obtained above with MD simulation
results. The idea is to measure $\langle r^2\rangle$ as a function of
time, extracting from it $\langle r^2\rangle_e$, $\tau$ and 
$\Omega$. $\langle r^2\rangle_e$ is obtained from the averaged value
around which $\langle r^2\rangle$ oscillates, $\tau$ by fitting the
relative maxima (or minima) to an exponential and $\Omega$ by
measuring the time between two relative maxima (or minima) that
can be separated by several periods.
The MD simulations are performed
as in section \ref{simulationsBoltzmann}. The only difference is that
the initial condition is generated in such a way that the initial
density profile is Gaussian. Hence, at the initial time the system is
closer to the breather and the condition for the theory to be valid,
i.e. that the distribution function has the same functional form that
the breather at the Boltzmann level, is reached faster. 

We have performed a series
of simulations with $N=1000$ disks and 
$q=0.2$ for different values of the
energy, in such a way that $\phi_0$ varies approximately in the interval $(0.005, 0.015)$. These values of the density are expected to be small enough for the theory to be valid, as the Boltzmann prediction describes accurately the dynamics in a wide enough time window. The results have been averaged over $50$ trajectories. With the
chosen values of the parameters, the system is close enough to
equilibrium so that the linear approximation is supposed to be
valid. This is supported by the fact that the non-linear solution of
Eqs. (\ref{sistM1}) and (\ref{sistM2}) with 
$\int d\bm{r}\text{Tr}\mathcal{P}^{(c)}(\bm{r},t)$ given by
Eqs. (\ref{ptc0}) and (\ref{ptc1}) is very close to the linear
approximation for the above mentioned parameters. For the considered values of the parameters, the minimum mean free path (the one at the origin) is of the same order as the size of the system measured by $\rho$, so that the conditions for the validity of hydrodynamics are doubtful. Nevertheless, this is not relevant in our case because, as it has been mentioned above, our predictions do not rely on a hydrodynamic description. In addition, this gives support to the analysis performed in \cite{gmgt24}, as the theoretical predictions do not depend on the regime.

\begin{figure}
\begin{center}
\includegraphics[angle=0,width=0.7\linewidth,clip]{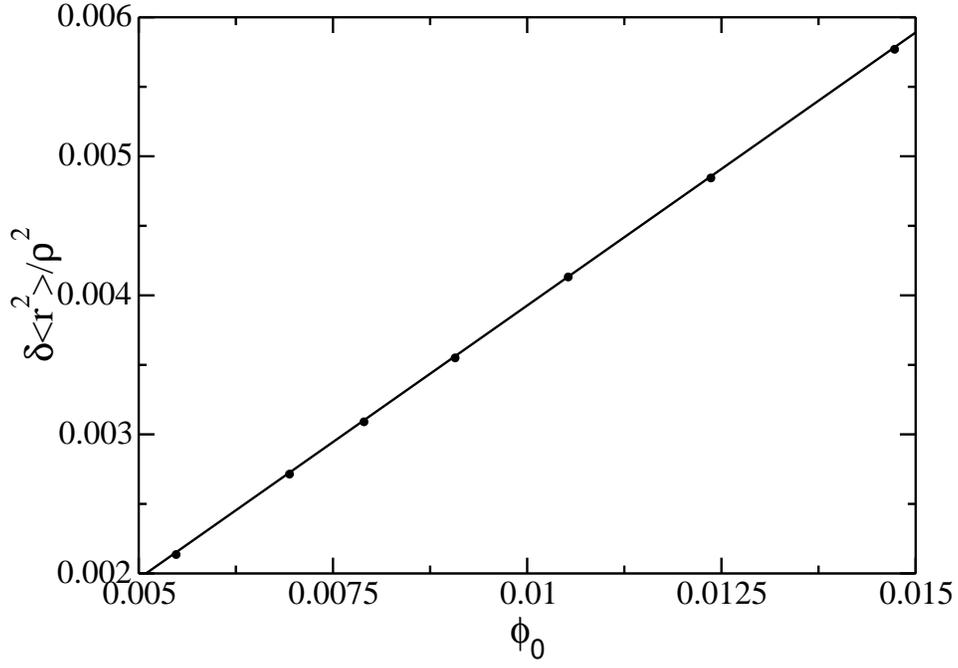}
\end{center}
\caption{$\frac{\delta\langle r^2\rangle_e}{\rho^2}$ as a function of
  the dimensionless density, $\phi_0$. The dots are the simulation
  results and the solid line is the theoretical prediction 
given by Eq. (\ref{deltar2}). }\label{deltar2_eq}
\end{figure}


\begin{figure}
\begin{center}
\includegraphics[angle=0,width=0.7\linewidth,clip]{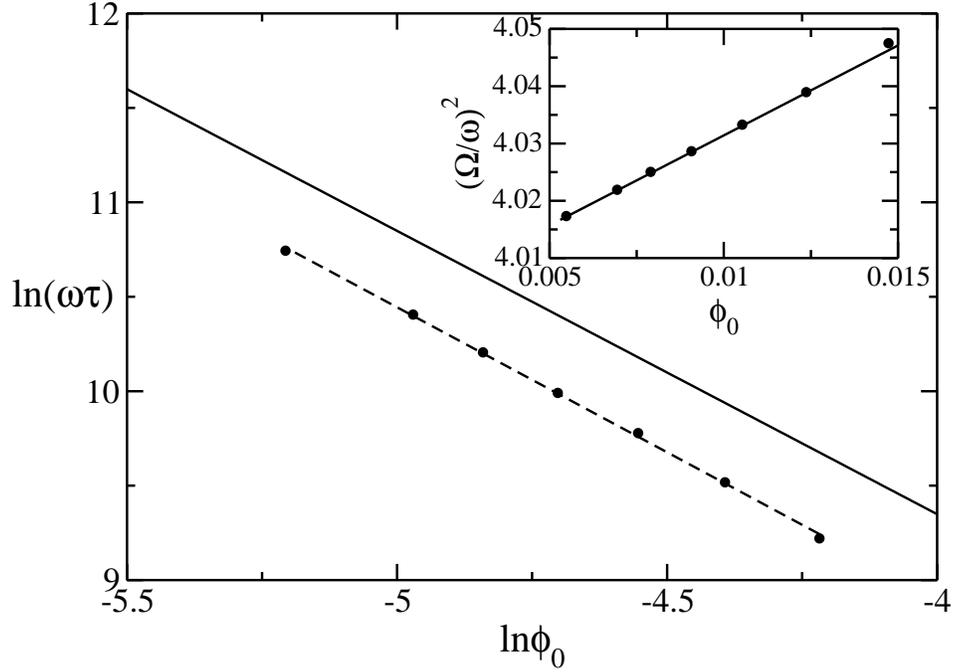}
\end{center}
\caption{$\omega{\tau}$ as a function of the
  dimensionless density, $\phi_0$, in logarithmic scale. The points are the simulation results, the 
  solid line the theoretical prediction given by
  Eq. (\ref{taud2}), and the dashed line is the linear fitting of the simulation results. In the inset, $\left(\frac{\Omega  }{\omega }\right)^2$ is plotted as a function of the
  dimensionless density, $\phi_0$. The dots are the simulation results and the solid line is the theoretical prediction given by Eq. (\ref{eqW})}.\label{tauFuncionEn}
\end{figure}

In Fig. \ref{deltar2_eq}, $\frac{\delta\langle r^2\rangle_e}{\rho^2}$ is
plotted as a function of the dimensionless density, $\phi_0$. The dots are the
simulation results and the solid line the theoretical prediction
given by Eq. (\ref{deltar2}). As can be seen, the agreement between
them is excellent. 
 
In Fig. \ref{tauFuncionEn}, $\omega\tau$ is plotted as a function
of $\phi_0$ in logarithmic scale. The dots are the
simulation results and the solid line the theoretical prediction given
by Eq. (\ref{eqTau}) that, for $d=2$ and expressed in terms of $N$ and $\phi_0$, takes the form
\begin{equation}\label{taud2}
\omega\tau=\frac{2\sqrt{2}}{\pi}\frac{N^{1/2}}{\phi_0^{3/2}}. 
\end{equation}
The dashed line is the linear fitting of the simulation results with slope $-1.53\pm 0.02$ in good agreement with Eq. (\ref{taud2}). 
As in the previous cases, the error bars
cannot be observed in the scale of the figure. Note that, in this
case, the agreement is not as good as in the previous two quantities. The quotient between the theoretical prediction and
the measured quantities is always of the order of $1.5$, indicating that, although the density dependence is perfectly captured by the theory, there are other not considered ingredients that renormalize the amplitude of $\phi_0^{-3/2}$. Similar results are obtained for other values
of the number of particles, so that the discrepancies are not due
to finite size effects. In the next
section we will further comment about that. 



\begin{figure}
\begin{center}
\includegraphics[angle=0,width=0.7\linewidth,clip]{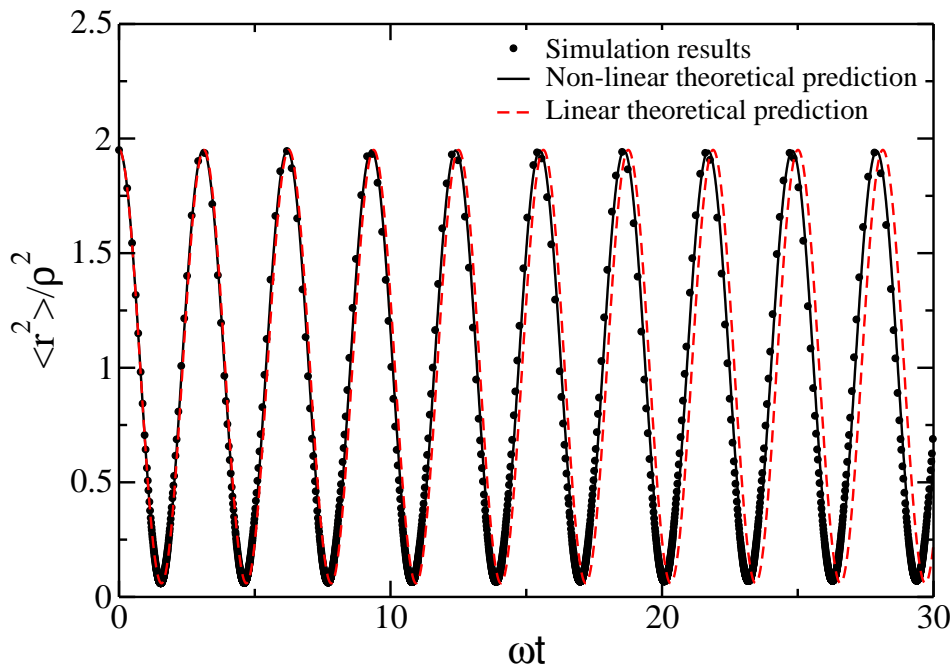}
\end{center}
\caption{$\frac{\langle r^2\rangle}{\rho^2}$ as a function of the dimensionless time, $\omega t$, for $\phi_0=0.013$ and $q=0.95$. The points are the MD simulation results, the (black) solid line is the non-linear theoretical prediction and the (red) dashed line is the linear theoretical prediction. The number of collisions per particle and per period of oscillation is around 10. }\label{ev1label}
\end{figure}

\begin{figure}
\begin{center}
\includegraphics[angle=0,width=0.7\linewidth,clip]{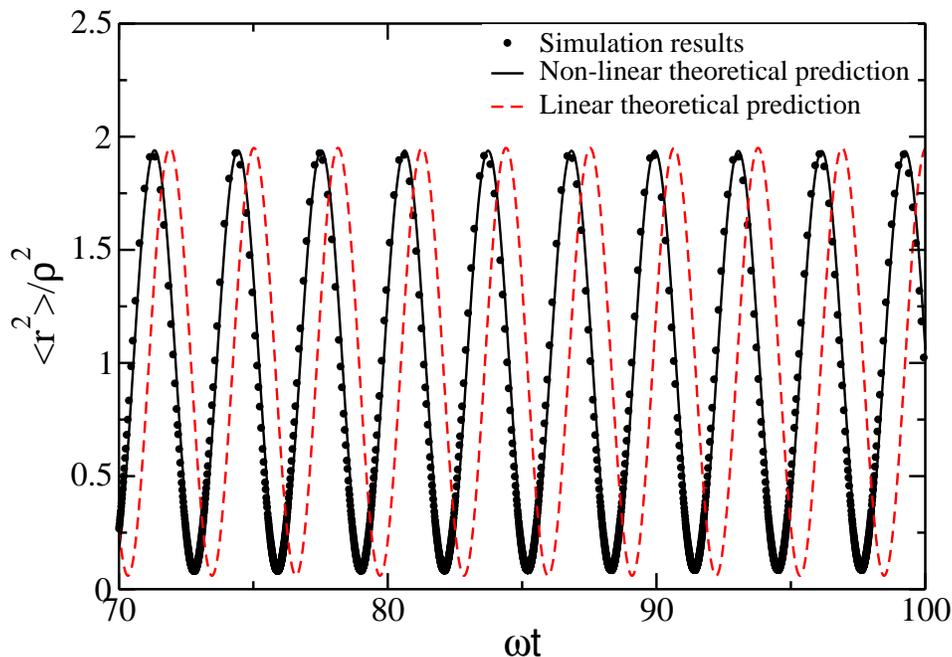}
\end{center}
\caption{ Same as in Fig. \ref{ev1label} but for a different time window. }\label{ev2label}
\end{figure}

\begin{figure}
\begin{center}
\includegraphics[angle=0,width=0.7\linewidth,clip]{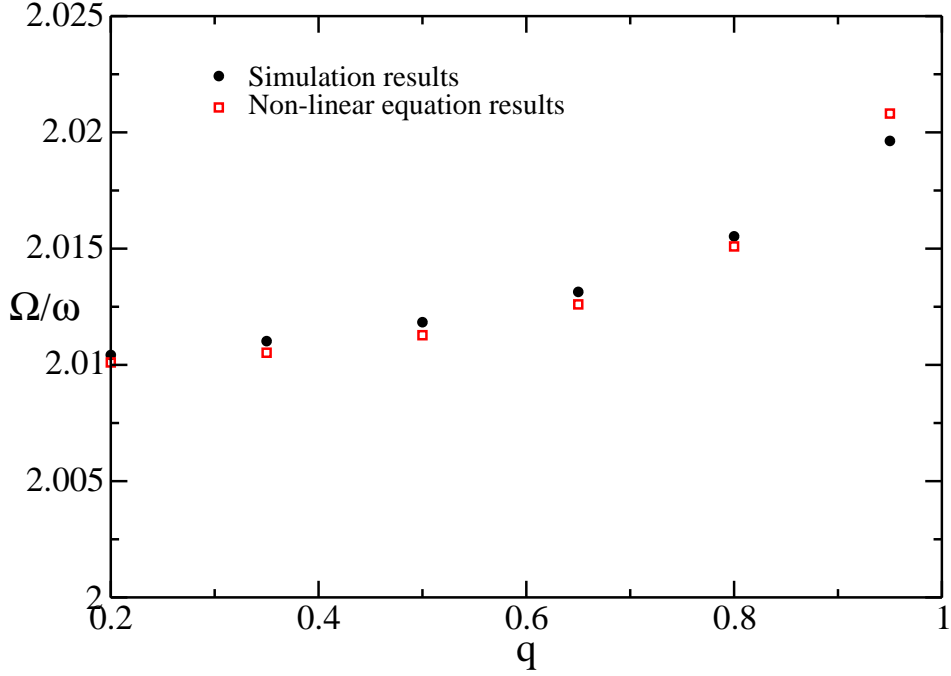}
\end{center}
\caption{$\frac{\Omega}{\omega}$ as a function of the relative amplitude, $q$, for $\phi_0=0.013$. The (black) circles are the simulation results and the (red) squares are the results extracted from the numerical solution of the non-linear equations.   }\label{omegalable}
\end{figure}

Finally, $\left(\frac{\Omega  }{\omega }\right)^2$ is
plotted as a function of the dimensionless density, $\phi_0$, in the inset of Fig. \ref{tauFuncionEn}. The dots are the
simulation results and the solid line the theoretical prediction given
by Eq. (\ref{eqW}).The agreement between the simulation 
results and the theoretical prediction is very good. The error bars in the measured
quantities cannot be observed in the scale of the figures.
We have also performed MD simulations very far from equilibrium where the linear approximation is supposed to fail. The values of the chosen parameters are $N=1000$, $\phi_0=0.013$ and the ``extreme'' value $q=0.95$, so that the amplitude of the oscillations of $\langle r^2\rangle$ is nearly $\rho^2$. In Fig. \ref{ev1label}, $\langle r^2\rangle$ is plotted as a function of the dimensionless time, $\omega t$, in the time window $\omega t \in [0,30]$. The points are MD simulation results, the (black) solid line is the non-linear theoretical prediction, while the (red) dashed line is the linear theoretical prediction. It can be seen that the agreement between the simulation results and the theoretical predictions is very good, although a small shift is found in the linear theoretical prediction for $\omega t>20$. This effect is amplified in Fig. \ref{ev2label}, where the same is plotted but for the time window $\omega t \in [70,100]$. Here, the linear theoretical prediction is clearly shifted with respect to the simulation results, while the non-linear one still perfectly fits the simulation data. More quantitatively, the values of the frequencies of the oscillations, $\frac{\Omega}{\omega}$, in the simulation results, non-linear theoretical prediction and linear theoretical predictions are $2.01963\pm 6\times 10^{-5}$, $2.02081$ and $2.01002$ respectively. We stress that the ``correction'' to the Boltzmann prediction in the non-linear case is of the order of twice the correction in the linear case. Note also that, for these times, the decay of the amplitude of the oscillations cannot be observed. The same analysis has been performed for different values of the relative amplitude, $q$, the results being plotted in Fig. \ref{omegalable}. The (black) circles are the simulation results and the (red) squares are the results extracted from the numerical solution of the non-linear equations. It is seen that the agreement between both is very good and that for $q=0.2$ the linear prediction is accurate. These results give a strong support to the validity of the non-linear equations, at least in the time window where $\int d\bm{r}\text{Tr}\mathcal{P}^{(c, 0)}(\bm{r},t)$ dominates the dynamics with respect $\int d\bm{r}\text{Tr}\mathcal{P}^{(c, 1)}(\bm{r},t)$. 


\section{Discussion and conclusions}
\label{sec:concl}

In this work, we have worked out the dynamics of a system of hard
particles that collide elastically and that are confined by an isotropic harmonic
potential: each particle of mass $m$ at point $\bm{r}$ is subject to a force $-m\omega^2 \bm{r}$. The study has been performed at two levels of description: at the
low-density limit where the Boltzmann equation describes the dynamics
of the system and beyond the low-density limit (although for low
densities). At the Boltzmann level, we have shown that, independently
of the initial condition, the system reaches a breather state in the
long-time limit. Concretely, if the system is studied in the frame of
reference where total momentum vanishes and assuming the total angular
momentum also vanishes, the breather is characterized by $N$ (number of particles), $\langle
r^2\rangle_0$, $\langle\bm{r}\cdot\bm{v}\rangle_0$ where $\langle...\rangle_0$ refers to an average in the initial condition, and the total energy per particle
$e$. This is a consequence of the fact that $\langle r^2\rangle$
verifies a closed equation, independently of the initial distribution function,
oscillating around the equilibrium value with frequency $2\,\omega$ (the
amplitude of the oscillations depends on $\langle r^2\rangle_0$ and
$\langle\bm{r}\cdot\bm{v}\rangle_0$). 
Equilibrium is only reached when $\langle r^2\rangle_0=\rho^2$ and $\langle\bm{r}\cdot\bm{v}\rangle_0$
vanishes. For low densities, MD simulation results are in excellent
agreement with the Boltzmann predictions. We have probed the above
mentioned prediction for $\langle r^2\rangle$ and we have shown that,
independently of the initial condition, after some collisions per
particle, the hydrodynamic fields reach the corresponding profiles,
oscillating with the corresponding frequency. Nevertheless, small
corrections are reported, eroding the amplitude of the oscillations very
slowly. The frequency of the oscillations and the mean value of
$\langle r^2\rangle$ are also slightly renormalized. All these effects
are due to the fact that a system of $N$ 
particles of finite diameter, $\sigma$, does never, strictly speaking, comply with the Boltzmann equation
level of description. 

The study of the dynamics of the system beyond the Boltzmann equation is
performed taking into account precisely this idea. By performing the
approximation $f_2(\bm{r}+\boldsymbol{\sigma}, \bm{v}_1,
\bm{r}, \bm{v},t)\approx 
f(\bm{r}+\boldsymbol{\sigma}, \bm{v}_1,t) f(\bm{r},
\bm{v},t)$ in the first equation of the BBGKY hierarchy, a closed
equation for the distribution function is obtained that takes into
account the finite size of the particles. Recall that the
Boltzmann equation is obtained with the additional approximation 
$f(\bm{r}+\boldsymbol{\sigma}, \bm{v},t)\approx f(\bm{r},
\bm{v},t)$. By taking moments in the dynamical equation,
evolution equations for $\langle r^2\rangle$ and
$\langle\bm{r}\cdot\bm{v}\rangle$ are obtained that, in
contrast to the Boltzmann ones, are not closed. A new term arises that
is proportional to the total collisional contribution to the
pressure, $\int
d\bm{r}\text{Tr}\mathcal{P}^{(c)}(\bm{r},t)$. Closed evolution
equations are obtained by assuming that the functional dependence of
the distribution function on  $\langle r^2\rangle$ and
$\langle\bm{r}\cdot\bm{v}\rangle$ is the same as the one in
the breather state at the Boltzmann level (this approximation is
supported by the previous MD simulation results) and by expanding the
distribution function to first order in the gradients. By performing a
linear analysis of the resulting equations, one finds that
$\langle r^2\rangle$ oscillates around the equilibrium value (that is
renormalized with respect to the Boltzmann value) with a given
frequency (that is also renormalized with respect to the Boltzmann
prediction), leading to a decay of the amplitude of the oscillations with a given 
relaxation time, $\tau$. While the renormalizations are due to the
zeroth order in the gradients expansion of $\int
d\bm{r}\text{Tr}\mathcal{P}^{(c)}(\bm{r},t)$, i.e. the excess
pressure, the decay of the oscillations is due to the first order
contribution. If a hydrodynamic description of the system is valid, this first order contribution can be understood as a fingerprint of the bulk
viscosity \cite{gmgt24}. In fact, the measure of $\tau$ can be considered to be a
direct probe of the bulk viscosity. We are not aware of any other process in which this transport coefficient appears in such a clear way. 

The agreement between the theoretical predictions and Molecular Dynamics
simulation results is excellent for the mean value and frequency of
the oscillations. For $\tau$, the agreement is decent, but it is not
as good as for the aforementioned quantities. This fact deserves some
comments. First, it must be mentioned
that the mismatch is not caused by the linearization process, as the numerical solution
of the complete non-linear equations leads to similar results. To analyze the problem, recall the main approximations made in the theory: a) Eq. (\ref{mcApp}), i.e. molecular chaos assumption ; b) Eq. (\ref{dfApp}), that approximates the distribution function to have the same functional form as in the breather state at the Boltzmann level with the hydrodynamic fields written in terms of $\langle r^2\rangle$ and $\langle \mathbf{r}\cdot\mathbf{v}\rangle$ ; c) the expansion in the gradients of the fields. Considering a), position correlations between the particles that are going to collide can be accounted for
by multiplying the collisional term by the pair correlation function at contact. This is the idea behind the Enskog equation \cite{e22} that incorporates position correlations of the particles at contact while neglecting velocity correlations. This would actually increase the collision
frequency and, hence, ``accelerate'' the process, obtaining a faster
decay, as desired. However, for the considered densities, the
corrections cannot reach the $1.5$ factor as the pair correlation function at contact is of the order of $1.02$, hence a 2\% correction. Although the density is very small, velocity correlations could contribute to the corresponding order in the density. However, their computation are complex and the analysis would require further investigations. Approximation c) could be improved by introducing more terms in the gradients expansion but it can hardly explain the discrepancy: as $\text{Tr}\mathcal{P}^{(c)}(\bm{r},t)$ is a scalar, the second
order in the gradients contribution is of the form 
$\text{Tr}\mathcal{P}^{(c,2)}(\bm{r},t)=c_1\nabla^2n
(\bm{r},t)+c_2[\nabla n
(\bm{r},t)]^2+c_3\sum_i\sum_j[\nabla_iu_j (\bm{r},t)]^2+
c_4\sum_i\sum_j\nabla_iu_j (\bm{r},t)\nabla_ju_i (\bm{r},t)$,
where $c_i$ with $i=1,\dots,4$ are Burnett coefficients, that do not
contribute to $\tau$ in the linear regime (in the non-linear case, 
a term of the form $(\dot{x})^2$ will appear that would not help,
either). Perhaps, the most plausible scenario to explain
the discrepancies is that approximation b) is not accurate enough at this order in the density. Within a hydrodynamic description, this could be improved by considering the hydrodynamic equations at the Enskog level. By linearizing them around the equilibrium
state, the relaxation time could be extracted, although it is difficult to go beyond the approximation performed in \cite{gmgt24} as the state
around which the linearization is performed is
position-dependent. Another possibility is to modify the ansatz given by Eq. (\ref{dfApp}) by incorporating more moments in the distribution function, e.g. all the possible position and velocity fourth moments. Taking into account the symmetry of the system, they are $\langle r^4\rangle$, $\langle r^2(\mathbf{r}\cdot\mathbf{v})\rangle$, $\langle r^2v^2\rangle$, $\langle (\mathbf{r}\cdot\mathbf{v})^2\rangle$, $\langle(\mathbf{r}\cdot\mathbf{v})v^2\rangle$ and $\langle v^4\rangle$. Work along these lines is in progress. 

The present work represents a first step in the study of the dynamics
of dense confined systems. A number of
interesting questions arise, pertaining to the behavior at high densities, or under anisotropic trapping. 
In the latter case, the decay to equilibrium features two contributions, one coming from
the anisotropy of the potential and one stemming from the finite
density. It would
be interesting to clarify the region of the parameters where one
effect or the other dominates. Work along these lines is also in
progress.

\acknowledgments

This research was supported by grant ProyExcel-00505 funded by Junta de Andaluc\'ia, and grant PID2021-126348NB-100 funded by
MCIN/AEI/10.13039/501100011033 and ERDF ``A way of making Europe'' and by the `Agence Nationale de la Recherche' grant No. ANR-18-CE30-0013.

\appendix

\section{Evolution equations for $\langle r^2\rangle$ and 
$\langle\bm{r}\cdot\bm{v}\rangle$}\label{app1}
We will proceed in two steps. We will first take velocity moments in
Eq. (\ref{Bbgky1}) to obtain balance equations for $n(\bm{r},t)$ and
$\bm{u}(\bm{r},t)$ and, then, we will take spatial moments
to obtain the desired equations. 

By integrating in the velocity in Eq. (\ref{Bbgky1}) and taking into
account that $\int d\bm{v}L[f_2]=0$ \cite{mgb18}, the density
equation is obtained
\begin{equation}\label{denEq}
\frac{\partial}{\partial t}n(\bm{r},t)
+\frac{\partial}{\partial\bm{r}}\cdot[n(\bm{r},t)
\bm{u}(\bm{r},t)]=0. 
\end{equation}
By multiplying by $\bm{v}$ in Eq. (\ref{Bbgky1}) and integrating
in the velocity space, one obtains
\begin{equation}\label{velEq}
\frac{\partial}{\partial t}[n(\bm{r},t)
\bm{u}(\bm{r},t)]+\int d\bm{v}\ \bm{v}\bm{v}
\cdot\frac{\partial}{\partial\bm{r}}f(\bm{r},\bm{v},t)
+\omega^2n(\bm{r},t)\bm{r}+\frac{1}{m}\frac{\partial}{\partial\bm{r}}
\cdot\mathcal{P}^{(c)}(\bm{r},t)=0, 
\end{equation}
where the collisional contribution to the pressure tensor has been
introduced through
\begin{equation}\label{colPreTenApp1}
\mathcal{P}_{ij}^{(c)}(\bm{r},t)=\frac{m}{2}\sigma^d\int
d\bm{v}_1\int d\bm{v}\int_0^1 d\lambda\int d\sig
\theta(-\bm{g}\cdot\sig)f_2\left[\bm{r}_1(\lambda, \sig),
\bm{v}_1, \bm{r}_2(\lambda, \sig), \bm{v}, t\right]
(\bm{g}\cdot\sig)^2\widehat{\sigma}_i\widehat{\sigma}_j, 
\end{equation}
with
\begin{equation}
\bm{r}_1(\lambda, \sig)=\bm{r}+\lambda\boldsymbol{\sigma},
\quad
\bm{r}_2(\lambda,
\sig)=\bm{r}+(\lambda-1)\boldsymbol{\sigma}, 
\end{equation}
and we have taken into account (see Ref. \cite{mgb18}) that 
\begin{equation}
\int d\bm{v}\
\bm{v}L[f_2]
=-\frac{1}{m}\frac{\partial}{\partial\bm{r}}
\cdot\mathcal{P}^{(c)}(\bm{r},t). 
\end{equation}

By multiplying by $r^2$ in Eq. (\ref{denEq}) and integrating in space,
Eq. (\ref{sistM1}) is obtained. By multiplying by $\bm{r}$ in
Eq. (\ref{velEq}) and integrating in space, one obtains 
\begin{equation}
\frac{d}{dt}\langle\bm{r}\cdot\bm{v}\rangle-\langle v^2\rangle
+\omega^2\langle r^2\rangle=\frac{1}{mN}\int
d\bm{r}\text{Tr}\mathcal{P}^{(c)}(\bm{r},t). 
\end{equation}
Taking into account that the energy is a constant of the motion,
$e=\frac{m}{2}\langle r^2\rangle+\frac{m}{2}\omega^2\langle r^2\rangle$,
Eq. (\ref{sistM2}) is obtained.  

\section{Gradient expansion of 
$\text{Tr}\mathcal{P}^{(c)}(\bm{r},t)$}\label{app2}

Taking into account molecular chaos, i.e. Eq. (\ref{mcApp}), 
into Eq. (\ref{colPreTen}), the expression for
$\text{Tr}\mathcal{P}^{(c)}$ is
\begin{equation}\label{eq1App2}
\text{Tr}\mathcal{P}^{(c)}(\bm{r},t)=\frac{m}{2}\sigma^d\int
d\bm{v}_1\int d\bm{v}\int d\lambda\int
d\sig\theta(\bm{g}\cdot\sig)
(\bm{g}\cdot\sig)^2f(\bm{r}-\lambda\boldsymbol{\sigma},
\bm{v}_1,t)
f[\bm{r}+(1-\lambda)\boldsymbol{\sigma}, \bm{v},t], 
\end{equation}
where we have also replaced $\sig$ by $-\sig$. By expanding the
integrand to first order in the gradients, it is
\begin{eqnarray}
f(\bm{r}-\lambda\boldsymbol{\sigma},
\bm{v}_1,t)
f[\bm{r}+(1-\lambda)\boldsymbol{\sigma}, \bm{v},t]\approx
f(\bm{r},\bm{v}_1,t)
f(\bm{r},\bm{v},t)\nonumber\\\label{eqExGra}
-\lambda f(\bm{r},\bm{v},t)\boldsymbol{\sigma}\cdot
\frac{\partial}{\partial\bm{r}}f(\bm{r},\bm{v}_1,t)
+(1-\lambda) f(\bm{r},\bm{v}_1,t)\boldsymbol{\sigma}\cdot
\frac{\partial}{\partial\bm{r}}f(\bm{r},\bm{v},t). 
\end{eqnarray}
And, by substituting Eq. (\ref{eqExGra}) into (\ref{eq1App2}), we have
\begin{equation}
\text{Tr}\mathcal{P}^{(c)}(\bm{r},t)=\text{Tr}\mathcal{P}^{(c,0)}(\bm{r},t)
+\text{Tr}\mathcal{P}^{(c,1)}(\bm{r},t), 
\end{equation}
with
\begin{eqnarray}
\text{Tr}\mathcal{P}^{(c,0)}(\bm{r},t)=\frac{m}{2}\sigma^d\int
d\bm{v}_1\int d\bm{v}\int
d\sig\theta(\bm{g}\cdot\sig)
(\bm{g}\cdot\sig)^2f(\bm{r},\bm{v}_1,t)f(\bm{r},
  \bm{v},t), \\
\text{Tr}\mathcal{P}^{(c,1)}(\bm{r},t)=\frac{m}{2}\sigma^d\int
d\bm{v}_1\int d\bm{v}\int
d\sig\theta(\bm{g}\cdot\sig)
(\bm{g}\cdot\sig)^2f(\bm{r},\bm{v}_1,t)\boldsymbol{\sigma}\cdot
\frac{\partial}{\partial\bm{r}}f(\bm{r},\bm{v},t), 
\end{eqnarray}
where the integral in $\lambda$ has been performed. 

Taking into
account that the distribution function has the form given by
Eq. (\ref{dfApp}), the zeroth order contribution is
\begin{eqnarray}
\text{Tr}\mathcal{P}^{(c,0)}(\bm{r},t)=\frac{m}{2}\sigma^d
n^2(\bm{r},t)\left[\frac{\beta(t)}{\pi}\right]^d\int
  d\bm{v}_1\int d\bm{v}\int d\sig\theta(\bm{g}\cdot\sig)
(\bm{g}\cdot\sig)^2e^{-\beta(t)[\bm{v}_1-\bm{u}(\bm{r},t)]^2 
-\beta(t)[\bm{v}-\bm{u}(\bm{r},t)]^2}\nonumber\\
=\frac{m n^2(\bm{r},t)\sigma^d}{2 \pi^{d}\beta(t)}\int d\bm{x}_1\int
  d\bm{x}_2
\int
  d\sig\theta(\bm{x_{12}}\cdot\sig)(\bm{x}_{12}\cdot\sig)^2e^{-x_1^2-x_2^2},
  \nonumber\\ 
\end{eqnarray}
where the new variables, 
$\bm{x}_1=\beta^{1/2}(t)[\bm{v}_1-\bm{u}(\bm{r},t)]$
and
$\bm{x}_2=\beta^{1/2}(t)[\bm{v}-\bm{u}(\bm{r},t)]$, 
have been introduced. The notation
$\bm{x}_{12}\equiv\bm{x}_1-\bm{x}_2$ is also
used. Finally, taking into account that
\begin{equation}
\int
  d\sig\theta(\bm{y}\cdot\sig)(\bm{y}\cdot\sig)^2
=\pi^{\frac{d-1}{2}}\frac{\Gamma\left(\frac{3}{2}\right)}
{\Gamma\left(\frac{d+2}{2}\right)}y^2, 
\end{equation}
and performing the Gaussian velocity integrals, 
one obtains 
\begin{equation}
\text{Tr}\mathcal{P}^{(c,0)}(\bm{r},t)=\frac{\pi^{d/2}}{\Gamma\left(\frac{d}{2}\right)}
n^2(\bm{r},t)\sigma^dT(t). 
\end{equation}

Taking into account that 
\begin{equation}
\int
  d\sig\theta(\bm{y}\cdot\sig)(\bm{y}\cdot\sig)^2 \widehat{\sigma}_i
=\frac{\pi^{\frac{d-1}{2}}}
{\Gamma\left(\frac{d+3}{2}\right)}yy_i, 
\end{equation}
the first order contribution is 
\begin{eqnarray}\label{traceP1}
\text{Tr}\mathcal{P}^{(c,1)}(\bm{r},t)=\frac{m\pi^{\frac{d-1}{2}}\sigma^{d+1}}
{2\Gamma\left(\frac{d+3}{2}\right)}\int d\bm{v}_1\int
  d\bm{v}
f(\bm{r},\bm{v}_1,t)g\bm{g}\cdot
\frac{\partial}{\partial\bm{r}}f(\bm{r},\bm{v},t). 
\end{eqnarray}
In addition, the gradient of the distribution function can be
explicitly written in the form
\begin{eqnarray}\label{gradientDist}
\frac{\partial}{\partial x_i}f(\bm{r},\bm{v},t)
=\frac{\partial}{\partial
  x_i}n(\bm{r},t)\left[\frac{\beta(t)}{\pi}\right]^{d/2}
e^{-\beta(t)[\bm{v}-\bm{u}(\bm{r},t)]^2}\nonumber\\
-\beta(t)
  n(\bm{r},t)
\left[\frac{\beta(t)}{\pi}\right]^{d/2}e^{-\beta(t)[\bm{v}-\bm{u}(\bm{r},t)]^2}
\left[2u_j(\bm{r},t)\frac{\partial}{\partial
  x_i}u_j(\bm{r},t)-2v_j \frac{\partial}{\partial
  x_i}u_j(\bm{r},t)\right], 
\end{eqnarray}
where the summation over repeated indexes has been used. 
By substituting Eq. (\ref{gradientDist}) into Eq. (\ref{traceP1}),
taking into account symmetry properties and performing the Gaussian
velocity integrals, the expression for $\text{Tr}\mathcal{P}^{(c,1)}$
of the main text is obtained, i.e. 
\begin{equation}
\text{Tr}\mathcal{P}^{(c,1)}(\bm{r},t)=-\frac{\sqrt{2}\pi^{\frac{d-1}{2}}}
{d\Gamma\left(\frac{d}{2}\right)}n^2(\bm{r},t)\sigma^{d+1}[2mT(t)]^{1/2}
\nabla\cdot\bm{u}(\bm{r},t). 
\end{equation}

\section{Evaluation of $\langle r^2\rangle_e$ in the first virial
  approximation }\label{app3}

The objective of this appendix is to calculate $\langle r^2\rangle_e$
in the first virial approximation. Let us assume that the state of the
system is described, in 
equilibrium, by the canonical $N$-particle density
\begin{equation}
\rho_N(\bm{r}_1,\bm{v}_1, \dots,
\bm{r}_N,\bm{v}_N)\propto
e^{-\frac{m}{2T}\sum_{i=1}^N(v_i^2+\omega^2r_i^2)}\Theta(\bm{R}), 
\end{equation}
where the notation $\bm{R}\equiv(\bm{r}_1, \dots,
\bm{r}_N)$ has been introduced and 
\begin{equation}
\Theta(\bm{R})=\prod_{i=1}^{N-1}\prod_{j>i}^N\theta(r_{ij}-\sigma), 
\end{equation}
where $r_{ij}\equiv\abs{\bm{r}_i-\bm{r}_j}$. In equilibrium,
we have 
\begin{equation}\label{r2eqVirial}
\langle r^2\rangle_e=\frac{\int
  d\bm{R}\Theta(\bm{R})r_1^2e^{-aR^2}}{\int
  d\bm{R}\Theta(\bm{R})e^{-aR^2}}, 
\end{equation}
where $a\equiv\frac{m\omega^2}{2T}$. 

Let us first evaluate the denominator of Eq. (\ref{r2eqVirial}). It
can be rewritten in the form
\begin{equation}\label{expansionDenominador}
\int
  d\bm{R}\Theta(\bm{R})e^{-aR^2}=\int
  d\bm{R}[\Theta(\bm{R})-1]e^{-aR^2}+
\int
  d\bm{R}e^{-aR^2}. 
\end{equation}
The second term can be calculated
\begin{equation}\label{denomParteTrivial}
\int
  d\bm{R}e^{-aR^2}=\left(\frac{\pi}{a}\right)^{\frac{dN}{2}}. 
\end{equation}
In order to evaluate the first term, it is assumed that the main
contribution comes from the volume in $\bm{R}$ for which only two
particles overlap
\begin{eqnarray}
\int
  d\bm{R}[\Theta(\bm{R})-1]e^{-aR^2}\approx\frac{N(N-1)}{2}\int
  d\bm{R}[\theta(r_{12}-\sigma)-1]e^{-aR^2}\nonumber\\\label{eqTeta}
 =\frac{N(N-1)}{2}\left(\frac{\pi}{a}\right)^{\frac{d}{2}(N-2)}\int 
  d\bm{r}_1\int
  d\bm{r}_2[\theta(r_{12}-\sigma)-1]e^{-a(r_1^2+r_2^2)}. 
\end{eqnarray}
Assuming that the exponential does not vary appreciably in distances of
the order of $\sigma$, by standard manipulations, we have
\begin{equation}\label{moco23}
\int d\bm{r}_1\int d\bm{r}_2
[\theta(r_{12}-\sigma)-1]e^{-a(r_1^2+r_2^2)}\approx
-\frac{\pi^d\sigma^d}
{\Gamma\left(\frac{d}{2}+1\right)(2a)^{d/2}}, 
\end{equation}
and, by substituting it into Eq. (\ref{eqTeta})
\begin{equation}\label{eqTeta2}
\int
  d\bm{R}[\Theta(\bm{R})-1]e^{-aR^2}\approx-\frac{N(N-1)}{2}
\left(\frac{\pi}{a}\right)^{\frac{d}{2}N}\left(\frac{a}{2}\right)^{d/2}
\frac{\sigma^d}{\Gamma\left(\frac{d}{2}+1\right)}. 
\end{equation}
Finally, taking into account Eqs. (\ref{denomParteTrivial}) and (\ref{eqTeta2}),
Eq. (\ref{expansionDenominador}) reads
\begin{equation}\label{denominadorFinal}
\int d\bm{R}\Theta(\bm{R})e^{-aR^2}\approx
\left(\frac{\pi}{a}\right)^{\frac{d}{2}N}
\left[1-\frac{N(N-1)}{2\Gamma\left(\frac{d}{2}+1\right)}
\left(\frac{a}{2}\right)^{d/2}\sigma^d\right].  
\end{equation}

To calculate the numerator of Eq. (\ref{r2eqVirial}), we proceed in a similar
fashion. 
\begin{equation}\label{expansionNumerador}
\int
  d\bm{R}\Theta(\bm{R})r_1^2e^{-aR^2}=\int
  d\bm{R}[\Theta(\bm{R})-1]r_1^2e^{-aR^2}+
\int
  d\bm{R}r_1^2e^{-aR^2}. 
\end{equation}
The second term reads
\begin{equation}\label{numeradorParteTrivial}
\int
  d\bm{R}r_1^2e^{-aR^2}=\left(\frac{\pi}{a}\right)^{\frac{dN}{2}}\frac{d}{2a}. 
\end{equation}
As above, in order to evaluate the first term, it is assumed that the
main contribution comes from the volume in $\bm{R}$ for which only
two particles overlap
\begin{eqnarray}\label{tetaR2Numer}
\int
  d\bm{R}[\Theta(\bm{R})-1]r_1^2e^{-aR^2}\approx
(N-1)\int d\bm{R}[\theta(r_{12}-\sigma)-1]r_1^2e^{-aR^2}\\
+\frac{(N-1)(N-2)}{2}\int
  d\bm{R}[\theta(r_{23}-\sigma)-1]r_1^2e^{-aR^2}. 
\end{eqnarray}
Again, assuming that the exponential does not vary appreciably in
distances of order $\sigma$, 
\begin{equation}\label{etatjkl}
\int d\bm{r}_1\int d\bm{r}_2
[\theta(r_{12}-\sigma)-1]r_1^2e^{-a(r_1^2+r_2^2)}\approx
-\frac{d\pi^d\sigma^d}{4a(2a)^{d/2}\Gamma\left(\frac{d}{2}+1\right)}. 
\end{equation}
Taking into account Eqs. (\ref{moco23}), (\ref{etatjkl}) and
performing the Gaussian integrals
\begin{eqnarray}\label{m145}
\int
  d\bm{R}[\Theta(\bm{R})-1]r_1^2e^{-aR^2}\approx
-(N-1)^2\frac{d\pi^{d/2}\sigma^d}{2^{\frac{d}{2}+2}a\Gamma\left(\frac{d}{2}+1\right)}
\left(\frac{\pi}{a}\right)^{\frac{d}{2}(N-1)}, 
\end{eqnarray}
and, then, with the aid of Eqs. (\ref{numeradorParteTrivial}) and
(\ref{m145}), Eq. (\ref{expansionNumerador}) reads 
\begin{eqnarray}\label{numeradorFinal}
\int
  d\bm{R}\Theta(\bm{R})r_1^2e^{-aR^2}\approx
\left(\frac{\pi}{a}\right)^{\frac{d}{2}N}\frac{d}{2a}
\left[1-(N-1)^2\frac{a^{d/2}\sigma^d}
{2^{\frac{d}{2}+1}\Gamma\left(\frac{d}{2}+1\right)}\right]. 
\end{eqnarray}

By substituting Eqs. (\ref{denominadorFinal}) and
(\ref{numeradorFinal}) into Eq. (\ref{r2eqVirial}), one obtains
\begin{eqnarray}
\langle r^2\rangle_e\approx\frac{d}{2a}
\left[1+N\frac{a^{d/2}\sigma^d}{2^{\frac{d}{2}+1}\Gamma\left(\frac{d}{2}+1\right)}\right]
\nonumber\\\label{r2EqFinalxx}
=\frac{dT}{m\omega^2}\left[1+N\frac{\sigma^d(m\omega^2)^{d/2}}{d2^d\Gamma(d/2)T^{d/2}}\right], 
\end{eqnarray}
where used have been done of the definition of $a$. Let us remark
that, here, $T$ is the actual temperature. In order to obtain the 
expression of the main text, it is necessary to express $T$ as a
function of $e$. Using 
$e=\frac{m}{2}\langle v^2\rangle_e+\frac{m}{2}\omega^2\langle r^2\rangle_e$
and the relation between the actual temperature and kinetic energy,
i.e. $dT=m\langle v^2\rangle_e$, one obtains (assuming that $e$ is
close to $dT$)
\begin{equation}\label{eqTaproxx}
dT\approx e\left[1-\frac{N\sigma^d}{d2^{\frac{d}{2}+1}\Gamma(d/2)}
\left(\frac{d}{2\rho^2}\right)^{d/2}\right], 
\end{equation}
where $\rho^2\equiv\frac{e}{m\omega^2}$. By substituting
Eq. (\ref{eqTaproxx}) into Eq. (\ref{r2EqFinalxx}), one finally
obtains
\begin{equation}
\langle r^2\rangle_e\approx\frac{e}{m\omega^2}\left[1
+\frac{N\sigma^d}{d2^{\frac{d}{2}+1}\Gamma(d/2)}
\left(\frac{d}{2\rho^2}\right)^{d/2}\right], 
\end{equation}
that coincides with the expression of the main text,
Eq. (\ref{deltar2}).


\begin{thebibliography}{27}
\bibitem [{not()}]{nota1}%
  \BibitemOpen
  \href@noop {} {}\bibinfo {note} {Strictly speaking the Boltzmann equation is
  obtained in the so-called Grad limit. For $N$ spheres of diameter $\sigma$
  the Grad limit is the double limit $N\to\infty$, $\sigma\to 0$ with
  $N\sigma^2$ finite. This limit implies $N\sigma^3\to 0$, which means that the
  total volume actually occupied by the particles goes to zero.}\BibitemShut
  {Stop}%
\end{thebibliography}
\end{document}